\documentclass{aa}
\usepackage[varg]{txfonts}
\usepackage{natbib}
\bibpunct{(}{)}{;}{a}{}{,}
\usepackage{graphicx}
\usepackage{caption}
\usepackage{subcaption}
\graphicspath{{./}{figures/}}
\usepackage{float}
\floatstyle{plaintop}
\restylefloat{table}
\usepackage{xcolor}
\usepackage[colorlinks=true,     linkcolor=blue, citecolor=blue, filecolor=blue, urlcolor=blue]{hyperref}

\begin{document}

    \title{Modelling carbon chain and complex organic molecules in the DR21(OH) clump}

   \author{P. Freeman
          \inst{1}
          \and
          S. Bottinelli 
          \inst{2}
          \and
          R. Plume
          \inst{1}
          \and
          E. Caux
          \inst{2}
          \and
          B. Mookerjea
          \inst{3}
          }

   \institute{Department of Physics and Astronomy, University of Calgary, Calgary, Canada\\ \email{pamela.freeman@ucalgary.ca}\\
         \and
             Institut de Recherche en Astrophysique et Plan\'etologie, Université de Toulouse, CNRS, UPS, CNES, Toulouse, France\\
        \and
            Department of Astronomy and Astrophysics, Tata Institute of Fundamental Research, Mumbai 400005, India
             }

\date{Submitted; Accepted}

\abstract{Star-forming regions host a large and evolving suite of molecular species. Molecular transition lines, particularly of complex molecules, can reveal the physical and dynamical environment of star formation.}
{We aim to study the large-scale structure and environment of high-mass star formation through single-dish observations of CH$_3$CCH, CH$_3$OH, and H$_2$CO.}{We have conducted a wide-band spectral survey with the Institut de radioastronomie millimétrique 30-m telescope and the 100-m Green Bank Telescope towards the high-mass star-forming region DR21(OH)/N44. We use a multi-component local thermodynamic equilibrium model to determine the large-scale physical environment near DR21(OH) and the surrounding dense clumps. We follow up with a radiative transfer code for CH$_3$OH to look at non-LTE behaviour. We then use a gas-grain chemical model to understand the formation routes of these molecules in their observed environments.}
{We disentangle multiple components of DR21(OH) in each of the three molecules. We find a warm and cold component each towards the dusty condensations MM1 and MM2, and a fifth broad, outflow component. We also reveal warm and cold components towards other dense clumps in our maps: N40, N36, N41, N38, and N48. We find thermal mechanisms are adequate to produce the observed abundances of H$_2$CO and CH$_3$CCH while non-thermal mechanisms are needed to produce CH$_3$OH.}{Through a combination of wide-band mapping observations, LTE and non-LTE model analysis, and chemical modeling, the chemical and physical environments of star-forming regions is revealed. This method allows us to disentangle the different velocity and temperature components within our clump-scale beam, a scale that links a star-forming core to its parent cloud. We find numerous warm, $\sim$20-80~K components corresponding to known cores and outflows in the region. We determine the production routes of these species to be dominated by grain chemistry.}


\maketitle

\section{Introduction}\label{section:intro}

The dynamic evolution of star formation leads to the production of a diverse suite of molecules. Molecular emission, then, reveals great detail about the physical and chemical conditions from which these species arise. In particular, wide-band mm and sub-mm observations are capable of capturing numerous rotational transition lines which are powerful diagnostics of their environment due to the different excitation conditions each line is sensitive to.

In early-stage protostellar sources in particular, when the accreting protostar is still enshrouded in an envelope, there are two categories of molecules distinguished for their diagnostic capabilities: complex organic molecules (COMs) and carbon-chain molecules (CCMs). These species are sensitive to their environment, and are valuable tracers alongside more common, simple species such as CO, HCN, or HCO$^+$. 

COMs, carbon-bearing saturated complex molecules, are abundant in the hot (T $>$ 100 K) and dense ($n_{\text{H}_2}$ $>$ 10$^6$ cm$^{-3}$) cores of star-forming regions (\citealt{blake1987, caselli1993, Cazaux2003, Ceccarelli2004}; see \citealt{herbst2009} for a review of these molecules). 

CCMs, on the other hand, are unsaturated hydrocarbons (e.g. C$_n$H, HC$_n$N) that are well-known in cold (T $\le$ 10 K) molecular clouds and starless cores (\citealt{avery1976, broten1978, kroto1978, little1978, sakai2010}). They are now also known to form in the warm gas enveloping protostellar systems in a process dubbed warm carbon chain chemistry (WCCC; \citealt{sakai2008, aikawa2008, sakai2013}).

COMs and CCMs, due to this chemical differentiation, can help classify protostellar systems. And while these species have been used to detail star-forming regions, especially hot cores, high-mass star-forming regions (HMSFR), which are capable of producing stars with M $>$ 8 M$_\odot$, are more difficult to study due to their relative rarity, clustered environments, and quick evolutionary timelines. As the evolutionary processes of HMSFR are not fully understood, observing their molecular emission is a valuable tool to study these regions. The first HMSFR found with evidence for WCCC was DR21(OH) \citep{mookerjea2012}, a nearby and prominent region within Cygnus X. 

DR21(OH) lies 2$\arcmin$ north of the DR21 H\,{\footnotesize II} region at a distance of 1.5~kpc \citep{rygl2012}. Its peak, known as DR21(OH)M or N44, has a total mass of about 10$^3$ M$_\odot$ and density of $n_{\text{H}_2}$ 10$^6$~cm$^{-3}$ and is further resolved into two sources, MM1 and MM2 (\citealt{woody1989, mangum1991, motte2007}). MM1 shows the stronger continuum emission at millimeter wavelengths \citep{liechti1997}. Both star-forming cores MM1 and MM2 are thought to be very young, with no visible H\,{\footnotesize II} region and only very weak continuum emission at the centimeter wavelengths \citep{argon2000}, i.e. in a pre-UCH\,{\footnotesize II} region phase. Dust continuum observations suggest that MM1 is the brighter source (L = 1.7$\times$10$^4$ L$_\odot$; consistent with a B0V star) and shows evidence of star formation, whereas MM2 (L = 1.2$\times$10$^3$ L$_\odot$; early B star), though more massive, is fainter and most likely at an even earlier stage of evolution. These sources have been further differentiated: \citet{zapata2012} found nine mm sources between MM1 (SMA 5-9) and MM2 (SMA 1-4) each of mass 8-24 M$_\odot$; \citet{minh2012} identified two hot subcores in mm observations of MM1 corresponding to SMA 6 and 7. Methanol masers (Class I, pumped collisionally) along a bipolar outflow contribute to the evidence of DR21(OH)M as a region of active star formation (\citealt{plambeck1990, kogan1998, kurtz2004, araya2009}).

Near DR21(OH)M are several other dense clumps: DR21(OH)S (or N48), DR21(OH)W (N38), DR21(OH)N1 (N41), DR21(OH)N2 (N40), and N36 (named by \citealt{mangum1991, chandler1993, motte2007}). These sources exist along the DR21 ridge, consisting of filamentary structures that influence collapse (\citealt{schneider2010, hennemann2012}). N48, N38, and N40 are massive IR-quiet cores noted as possible precursors to high-mass protostars — in \citet{motte2007} the lack of IR emission suggests they are not forming a star yet, but SiO emission is indicative an outflow is being powered by a protostellar object. \cite{mangum1991} find N48 and N38 as high luminosity objects, consistent with B-type stars, but they also do not detect any maser emission which suggests these sources would be in the early stages of star formation. In total, these sources make up a parsec-scale region with a mass of 4900 M$_\odot$ (based on 1.2 mm continuum emission, \cite{schneider2010}).

Interferometric observations have catalogued DR21(OH)M as a rich molecular source including hydrocarbons and COMs (\citealt{zapata2012, minh2012, girart2013, OrozcoAguilera2019}), while observations of the larger scale environment detail the continuum emission and that of simple species such as CO, CS, SiO, N$^2$H$^+$, and HCO$^+$ (\citealt{richardson1994, lai2003, motte2007, schneider2010}). We aim to link the molecular environment of DR21(OH)M at core scales to the larger clump and filament it resides in. We will use single-dish observations to map CH$_3$CCH as a CCM tracer as well as CH$_3$OH and H$_2$CO as COM tracers, and use an LTE model to determine the physical parameters of different gaseous components within the DR21(OH) region. The rest of the paper is structured as follows: Section~\ref{section:observations} shows the observations and the data set used, Sect.~\ref{section:ltemodel} includes the local thermodynamic equilibrium (LTE) modelling results and analysis, Sect.~\ref{section:discussion} presents a chemical evolution model and discusses the findings, and Sect.~\ref{section:summary} provides a brief summary.

\section{Observational data}
\label{section:observations}

\subsection{Observations}

We observed DR21(OH) in multiple frequency ranges with the IRAM 30-m telescope and the 100-m Green Bank Telescope (GBT). A detailed summary of these observations is found in~\citet{freeman2023}.

IRAM 30-m observations were completed November 2020 and April 2021 (project codes 021-20 and 122-20, PIs S. Bottinelli and R. Plume). Using the on-the-fly observational mode with position switching, we mapped a $1' \times 1.5'$ region in DR21(OH) simultaneously in the frequency ranges 131.2 - 138.98~GHz and 287.22 - 295~GHz. The offset position was -120$\arcsec$ horizontally and -240$\arcsec$ vertically. The map center is $\alpha$J(2000)$=20^\mathrm{h}39^\mathrm{m}01\fs 00$ and 
$\delta$J(2000)$=+42\degr 22\arcmin48\farcs0$. Data reduction, and the production of maps, was completed with GILDAS/CLASS\footnote{https://www.iram.fr/IRAMFR/GILDAS}. At 291~GHz, the spectral channel width is 781~kHz or 0.80~km~s$^{-1}$, the beam size is 9.3$\arcsec$, and the pixel size is 4.4$\arcsec$. At 135~GHz, the spectral channel width is 391~kHz or 0.88~km~s$^{-1}$, the beam size is 20.5$\arcsec$, and the pixel size is 9.7$\arcsec$. Given the larger beam size, the 135 GHz data will not be discussed in this paper.

GBT observations were completed in March 2021, in the frequency ranges 84.5-85.75~GHz and 95.55-96.8~GHz using all 16 beams of the ARGUS focal plane array and the VErsatile GBT Astronomical Spectrometer (VEGAS) spectral line backend (project code 21A-039, PI P. Freeman). We obtained 1$\arcmin$ DAISY on-the-fly maps for both frequency ranges, centered as above, utilizing position-switching for the reference measurements. The spectral sampling with spectrometer mode 2 is 92~kHz in both frequency ranges, or 0.32~km s$^{-1}$ at 85~GHz and 0.28~km s$^{-1}$ at 96~GHz. The beam size is 9.2$\arcsec$ at 96~GHz and 10.0$\arcsec$ at 85~GHz, both with a pixel size of 2.0$\arcsec$. The data were reduced and calibrated using GBTIDL\footnote{https://gbtidl.nrao.edu/index.shtml}. In Sect.~\ref{section:ltemodel}, the data are presented in units of $T_\mathrm{A}^*$. The LTE model, computed by definition in $T_\mathrm{mb}$ is converted to $T_\mathrm{A}^*$ using the telescope $B_\mathrm{eff}$/$F_\mathrm{eff}$ values provided by the respective observatories: GBT at 85 GHz, 0.4545; GBT at 96 GHz, 0.3838; IRAM EMIR at 290 GHz, 0.547.\\

\subsection{Species of study}

In these data sets we detect numerous lines of CH$_3$CCH, a CCM tracer, and of CH$_3$OH and H$_2$CO, to be used as COM tracers. The identification of these lines is presented in Freeman et al. (submitted), and was done with CASSIS\footnote{https://cassis.irap.omp.eu/}~\citep{vastel2015} using the Cologne Database for Molecular Spectroscopy\footnote{https://cdms.astro.uni-koeln.de/classic/} (CDMS) catalogue~\citep{MULLER2005}.

\begin{table*}
\caption{Rotational transition lines of selected molecules in the observed frequency ranges, from \cite{freeman2023}.}
\label{tab:lines}
\centering
\begin{tabular}{ l l l c c c l l c c c }
\hline\hline
Molecule & \multicolumn{2}{c}{Quantum Numbers} & Frequency & $E_{\rm up}$ &  $A_{\rm ij}$ & \multicolumn{2}{c}{Quantum Numbers} & Frequency &
$E_{\rm up}$ & $A_{\rm ij}$ \\
 & \multicolumn{2}{c}{} & [GHz] & [K] & [10$^{-5}$ s$^{-1}$] & \multicolumn{2}{c}{} & [GHz] & [K] & [10$^{-5}$ s$^{-1}$]
\\
\hline
CH$_3$CCH & 5$_0$-$4_0$ & {e} & 85.4573 & 12.31 & 0.2
            & 17$_0$-17$_0$ & {e} & 290.5020 & 125.50 & 8.5\\
             & 5$_1$-4$_1$ & {e} & 85.4557 & 19.53 & 0.2
             & 17$_1$-17$_1$ & {e} & 290.4965 & 132.73 & 8.5\\
             & 5$_2$-4$_2$ & {e} & 85.4508 & 41.21 & 0.2
             & 17$_2$-17$_2$ & {e} & 290.4799 & 154.40 & 8.4\\
             & 5$_3$-4$_3$ & {a} & 85.4426 & 77.34 & 0.1
             & 17$_3$-17$_3$ & {a} & 290.4522 & 190.52 & 8.3 \\
            \hline
CH$_3$OH  & 2$_{0,2,0}$-1$_{0,1,0}$ & {A+} & 96.7413 & 6.96 & 0.3 
          & 4$_{3,2,1}$-5$_{2,4,1}$ & {E} & 288.7056 & 70.93 & 1.8\\
          & 2$_{1,2,2}$-1$_{1,1,2}$ & {E} & 96.7394 & 12.54 & 0.3 & 6$_{2,5,1}$-5$_{2,4,1}$ & {E} & 290.3077 & 71.01 & 9.3\\
          & 2$_{0,2,1}$-1$_{0,1,1}$ & {E} & 96.7445 & 20.09 & 0.3 & 6$_{2,4,2}$-5$_{2,3,2}$ & {E} & 290.3073 & 74.66 & 9.5\\
          & 2$_{1,2,0}$-1$_{1,1,0}$ & {A+} & 95.9143 & 21.45 & 0.2 & 6$_{2,5,0}$-5$_{2,4,0}$ & {A-} & 290.1847 & 86.46 & 9.5\\
          & 2$_{1,1,1}$-1$_{1,0,1}$ & {E} & 96.7555 & 28.01 & 0.3 & 6$_{2,4,0}$-5$_{2,3,0}$ & {A+} & 290.2641 & 86.47 & 9.5\\
          & 5$_{1,5,2}$-4$_{0,4,1}$ & {E} & 84.5212 & 40.39 & 0.2 & 6$_{3,4,1}$-5$_{3,3,1}$ & {E} & 290.2132 & 96.47 & 8.0\\
          & 6$_{0,6,0}$-5$_{0,5,0}$ & {A+} & 290.1106 & 48.74 & 10.6 & 6$_{3,4,0}$-5$_{3,3,0}$ & {A+} & 290.1895 & 98.55 & 7.9\\
          & 3$_{2,1,0}$-4$_{1,4,0}$ & {A+} & 293.4641 & 51.64 & 2.9 & 6$_{3,3,0}$-5$_{3,2,0}$ & {A-} & 290.1905 & 98.55 & 7.9\\
          & 6$_{1,6,2}$-5$_{1,5,2}$ & {E} & 290.0697 & 54.32 & 10.3 & 6$_{3,3,2}$-5$_{3,2,2}$ & {E} & 290.2097 & 111.47 & 8.0\\
          & 6$_{0,6,1}$-5$_{0,5,1}$ & {E} & 289.9394 & 61.86 & 10.6& 6$_{4,3,0}$-5$_{4,2,0}$ & {A-} & 290.1613 & 129.10 & 5.9\\
          & 6$_{1,6,0}$-5$_{1,5,0}$ & {A+} & 287.6708 & 62.87 & 10.1 & 6$_{4,3,0}$-5$_{4,2,0}$ & {A+} & 290.1614 & 129.10 & 5.9\\
          & 6$_{1,5,0}$-5$_{1,4,0}$ & {A-} & 292.6729 & 63.71 & 10.6 & 6$_{4,3,2}$-5$_{4,2,2}$ & {E} & 290.1624 & 136.66 & 5.9\\
          & 6$_{1,5,1}$-5$_{1,4,1}$ & {E} & 290.2487 & 69.81 & 10.6 & 6$_{4,2,1}$-5$_{4,1,1}$ & {E} & 290.1833 & 144.76 & 5.9\\
          \hline
H$_2$CO     & 4$_{0,4}$-3$_{0,3}$ & {p} & 290.6234 & 34.90 & 69
           & 4$_{3,2}$-3$_{3,1}$ & {o} & 291.3804 & 140.94 & 30.4\\
            & 4$_{2,3}$-3$_{2,2}$ & {p}  & 291.2378 & 82.07 & 52.1
            & 4$_{3,1}$-3$_{3,0}$ & {o} & 291.3844 & 140.94 & 30.4\\
            & 4$_{2,2}$-3$_{2,1}$ & {p} & 291.9481 & 82.12 & 52.5
            & & & & & \\
\hline
\end{tabular}
\tablefoot{The line properties are taken from the CDMS and VASTEL catalogues. Limits were set for E$_{\mathrm{up}}$ $<$ 150 K (200 K for CH$_3$CCH) and $A_{\mathrm{ij}}$ $>$ 1.0$\times 10^{-6}$ (2.0 $\times 10^{-6}$ for CH$_3$OH).}
\end{table*}

Propyne, or methyl acetylene, CH$_3$CCH, is a symmetric rotor with many transitions closely spaced in frequency. Numerous lines are thus captured in a small bandwidth, and are useful for analysis of the physical conditions, especially temperature, in LTE (\citealt{irvine1981, Askne1984, kuiper1984, Bergin1994}). It has long been detected in low- and high-mass star-forming regions and is used as a tracer of chemical complexity (\citealt{snyder1973, lovas1976, Cazaux2003, taniguchi2018b, gianetti2017, santos2022}).

Methanol, CH$_3$OH, is the simplest alcohol molecule and an asymmetric top molecule \citep{ball1970}. It has numerous transitions in the mm and sub-mm range which are abundant and commonly used as probes of the physical environment. H$_2$CO was the first polyatomic organic molecule detected in space \citep{Snyder1969}, and while not technically a complex species it is chemically associated with larger organic molecules. Both H$_2$CO and CH$_3$OH are formed by the successive hydrogenation of CO on dust grain surfaces, and are precursor molecules to larger COMs (\citealt{Charnley1997, garrod2013}). These two species are detected across interstellar space: in cold clouds, hot cores, outflows, shocks, the Galactic centre, and external galaxies (see \citealt{herbst2009} and references therein).

\begin{figure*}[h!]
\centering
\includegraphics[scale=0.6]{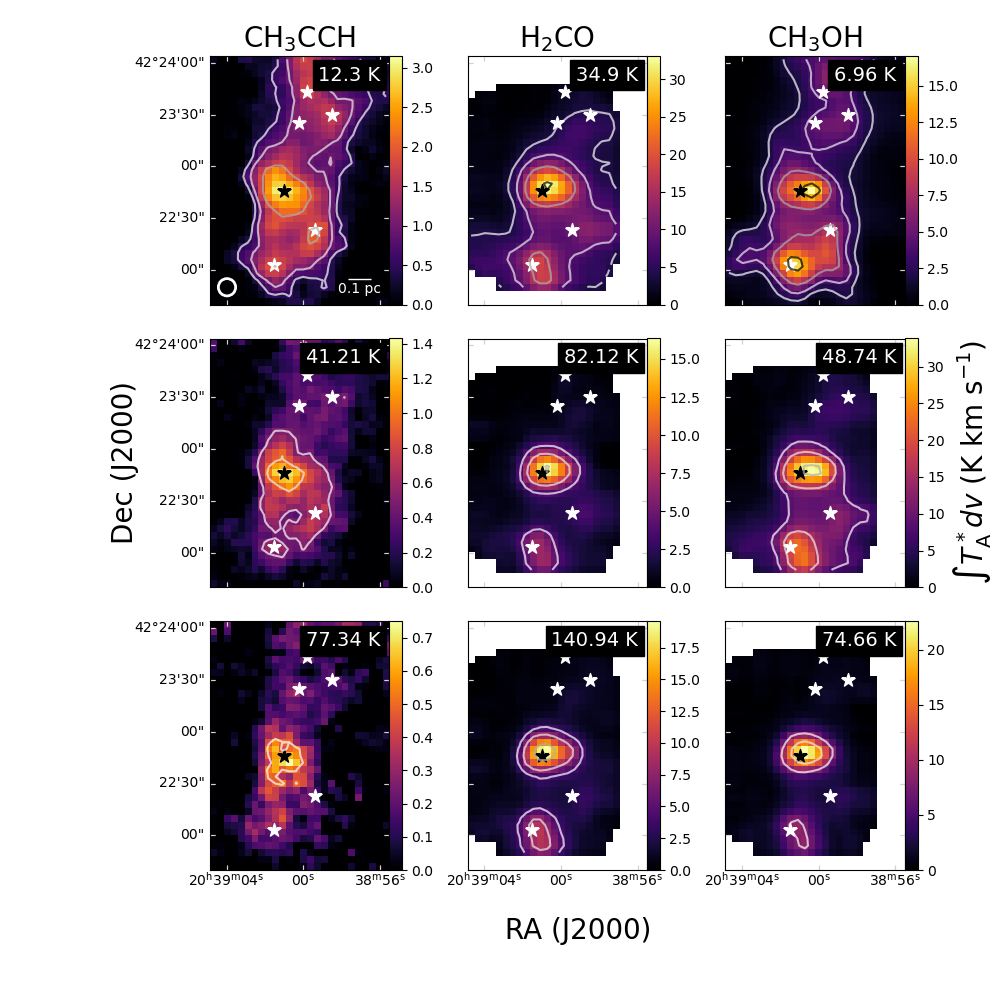}
\caption{Integrated intensity maps for select transition lines of CH$_3$CCH, H$_2$CO, and CH$_3$OH. The dense clumps as identified in \cite{motte2007} are shown in stars, with the main clump N44 in black (see also Fig.~\ref{fig:dr21regions}).}
\label{fig:intintensity}
\end{figure*}

Table~\ref{tab:lines} lists the transition lines seen in our frequency ranges, with certain conditions based on the sensitivity of our observations. We limit the scope of lines: for CH$_3$CCH an $A_{\mathrm{ij}}$ $>$ 1$\times10^{-6}$ s$^{-1}$ and E$_{\mathrm{up}}$ $<$ 200~K (to account for the a-CH$_3$CCH line that has E$_{\mathrm{up}}$ $<$ 150~K, but has a higher tabulated value when the a- and e- types are combined in CDMS); for H$_2$CO an $A_{\mathrm{ij}}$ $>$ 1$\times10^{-6}$ s$^{-1}$ and E$_{\mathrm{up}}$ $<$ 150 K; for CH$_3$OH an $A_{\mathrm{ij}}$ $>$ 2$\times 10^{-6}$ s$^{-1}$ (to remove the 84.5~GHz maser line) and E$_{\mathrm{up}}$ $<$ 150 K. Integrated intensity maps of select lines covering a range of upper energy levels are displayed in Figure~\ref{fig:intintensity}.

Class I masers are often found in young star-forming regions (see \citealt{Ladeyschikov2019}), including in DR21(OH) (\citealt{slysh1997, araya2009}). In particular, the CH$_3$OH line at 84.5 GHz is a known maser in DR21(OH) (\citealt{batrla1988}). These masers, which are collisionally pumped, are unlikely to be well reproduced by LTE. Thus, to be careful with our LTE analysis, we removed this line.

\section{LTE modelling -- results and analysis}
\label{section:ltemodel}

An LTE model was used to determine the physical parameters of the observed region -- size, line width, column density, excitation temperature, and source velocity. In LTE, we assume the excitation temperature represents the gas temperature. We use a python-based LTE model based on the CASSIS formalism\footnote{http://cassis.irap.omp.eu/docs/RadiativeTransfer.pdf} (previously described in \citealt{freeman2023} for a similar analysis of AFGL 2591 and IRAS 20126+4104). The LTE model identifies all molecular transitions for one or more species and simultaneously fits these lines defined by the physical parameters. The resulting fit is the combination of parameters that best reproduces the spectral line profile as determined by a least-squares minimisation (represented as a reduced $\chi^2$ value). In the model, we use the Cologne Database for Molecular Spectroscopy\footnote{https://cdms.astro.uni-koeln.de/classic/} (CDMS; \citealt{MULLER2005}) tags for each species. The species types --  e.g. a- and e-CH$_3$CCH, o- and p-$c$-C$_3$H$_2$, A- and E-CH$_3$OH, o- and p-H$_2$CO -- were not differentiated as we do not detect enough lines of each.

In the simplest case, an observed spectral line will arise from one physical gas component modeled by a single Gaussian profile with a centroid velocity ($v_\mathrm{LSR}$), line width ($\Delta v_\mathrm{FWHM}$), and brightness ($T_\mathrm{A}$*). However, in the case where the telescope's beam encompasses multiple physical components, the observed spectra may have to be modeled using multiple Gaussian profiles.

\begin{figure*}[h!]
\centering
\includegraphics[scale=0.6]{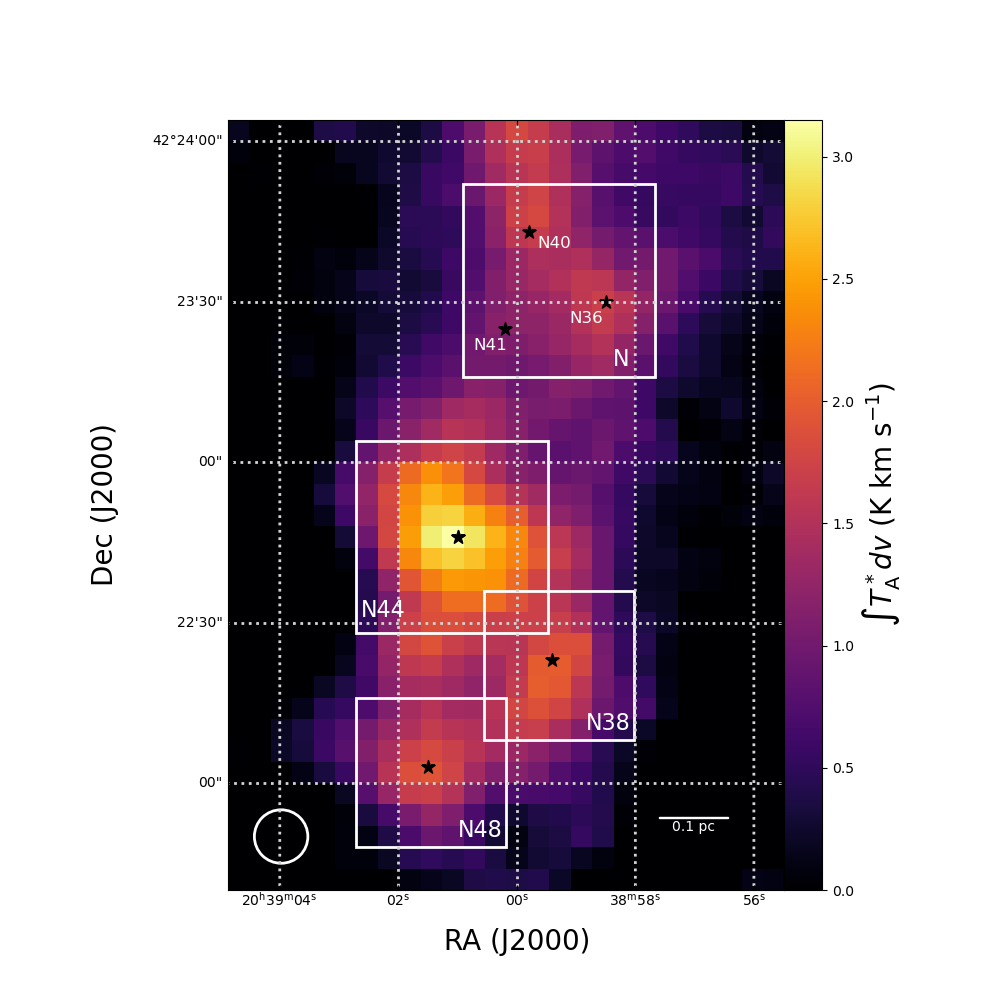}
\caption{The regions isolated for LTE modelling near and in DR21(OH), marked in white boxes over an integrated intensity map of the 12~K CH$_3$CCH $5_0$-$4_0$ line. From the top, the regions include: N40, N36, and N41 (labelled N); N44; N38; N48. The dense clumps, as designated by \cite{motte2007}, are marked by stars.}
\label{fig:dr21regions}
\end{figure*}

We treated each species, CH$_3$CCH, CH$_3$OH, and H$_2$CO, separately in our analysis. To determine the components of each species, we started with a one-component fit and looped the model over all pixels in the map. Where a one-component fit was not adequate, either by a visual inspection of the spectral line profile or by the resulting $\chi^2$ analysis, we added components as needed. This divided the map into regions described by their own, distinct, physical components. The regions, marked in Fig.~\ref{fig:dr21regions}, correspond to the clumps found in \cite{motte2007}.  We gained an idea of what components were necessary based on (a) taking an example spectrum and fitting multiple Gaussians in CASSIS to find velocity and line width information, and (b) knowledge from the literature on what velocity and temperature components we expect to see at this scale. The results for these fits are presented in Table~\ref{tab:ltemodel}.

\begin{table*}
\caption{LTE model results. Values without errors were fixed in the model.}
\label{tab:ltemodel}
\centering
\begin{tabular}{ l | l | c c c c  }
\hline\hline
Species & Component & $N_{\mathrm{tot}}$ & $T_{\mathrm{ex}}$ & FWHM & $v_{\mathrm{lsr}}$ \\
&  & (cm$^{-2}$) & (K) & (km s$^{-1}$) & (km s$^{-1}$) \\
\hline
CH$_3$CCH & N44 1 & 1.04 ($\pm$0.04) $\times$ $10^{15}$ & 42.2 ($\pm$1.4) & 2.8 & $-$4.09 ($\pm$0.04) \\
 & N44 2 & 7.11 ($\pm$0.37) $\times$ $10^{14}$ & 38.2 ($\pm$1.9) & 2.8 & $-$0.9  \\
  & N 1 & 5.24 ($\pm$0.35) $\times$ 10$^{14}$ & 30.0 & 2.6 & $-$3.19 ($\pm$0.07)  \\
  & N 2 & 3.02 ($\pm$0.14) $\times$ 10$^{14}$  & 25.0 & 1.2 &  $-$3.8 \\
 & N38 &  7.93 ($\pm$0.32) $\times$ $10^{14}$ & 28.6 ($\pm$1.3) & 2.9 ($\pm$0.1) &  $-$3.13 ($\pm$0.04) \\
 & N48 & 7.79 ($\pm$0.32) $\times$ $10^{14}$  & 34.1 ($\pm$1.4) & 2.5 ($\pm$0.1) &  $-$3.95 ($\pm$0.04) \\
            \hline
CH$_3$OH &  N44 1 & 2.15 ($\pm$0.13) $\times$ $10^{15}$ & 82.0 & 4.0 & $-$3.57 ($\pm$0.13) \\
 & N44 2 & 1.85 ($\pm$0.13) $\times$ $10^{15}$ & 75.0 & 4.0 & $-$0.5 \\
& N44 3 & 2.42 ($\pm$0.06) $\times$ $10^{15}$ & 22.0 & 2.8 & $-$4.2   \\
 & N44 4 & 2.16 ($\pm$0.07) $\times$ $10^{15}$  & 15.0 & 2.8 & $-$0.9  \\
  & N44 5 & 1.65 ($\pm$0.08) $\times$ $10^{15}$  & 25.0 & 15.0 & $-$8.0  \\
  & N 1 &  3.86 ($\pm$0.32) $\times$ $10^{14}$ & 30.0 & 3.6 & $-$2.50 ($\pm$0.03)  \\
  & N 2 & 1.02 ($\pm$0.11) $\times$ $10^{15}$ & 6.0 & 3.0 &  $-$3.41 ($\pm$0.06) \\
 & N 3 & 3.05 ($\pm$0.31) $\times$ $10^{14}$ & 6.0 & 1.2 & $-$3.17 ($\pm$0.03)  \\
 & N38 1 & 8.29 ($\pm$0.41) $\times$ $10^{14}$ & 30.0 & 3.6 & $-$2.20 ($\pm$0.09) \\
 & N38 2 & 1.22 ($\pm$0.07) $\times$ $10^{15}$ & 8.0 & 3.2 & $-$3.4 \\
  & N38 3 & 4.51 ($\pm$0.49) $\times$ $10^{14}$  & 8.0 & 1.8 & $-$1.75 ($\pm$0.08) \\
 & N48 1 & 2.91 ($\pm$0.10) $\times$ 10$^{15}$ & 20.0 & 3.1 & $-$4.04 ($\pm$0.02) \\
 & N48 2 & 9.78 ($\pm$1.51) $\times$ 10$^{14}$ & 8.0 & 3.2 & $-$3.78 ($\pm$0.03) \\
 \hline
H$_2$CO & N44 1 & 8.02 ($\pm$0.21) $\times$ $10^{14}$ & 80.5 ($\pm$3.0) & 4.0 & $-$4.11 ($\pm$0.04)  \\
 & N44 2 & 5.03 ($\pm$0.19) $\times$ $10^{14}$ & 83.9 ($\pm$4.3) & 4.0 & $-$0.5 \\
 & N44 3 & 2.33 ($\pm$0.12) $\times$ $10^{14}$ & 80.0 & 15 & $-$8.0 \\
  & N & 1.45 ($\pm$0.04) $\times$ $10^{14}$  & 51.3 ($\pm$2.1) & 4.1 ($\pm$0.1) &  $-$2.11 ($\pm$0.06) \\
 & N38 & 2.81 ($\pm$0.06) $\times$ $10^{14}$  & 52.5 ($\pm$1.4) & 3.7 ($\pm$0.1) & $-$2.31 ($\pm$0.04)  \\
 & N48 & 5.40 ($\pm$0.12) $\times$ $10^{14}$  & 67.6 ($\pm$2.4) & 3.7 ($\pm$0.1) & $-$4.08 ($\pm$0.04)  \\
\hline
\end{tabular}
\end{table*}

\subsection{N44}
\label{sect:n44}
\begin{figure*}[h!]
\centering
\includegraphics[width=\textwidth]{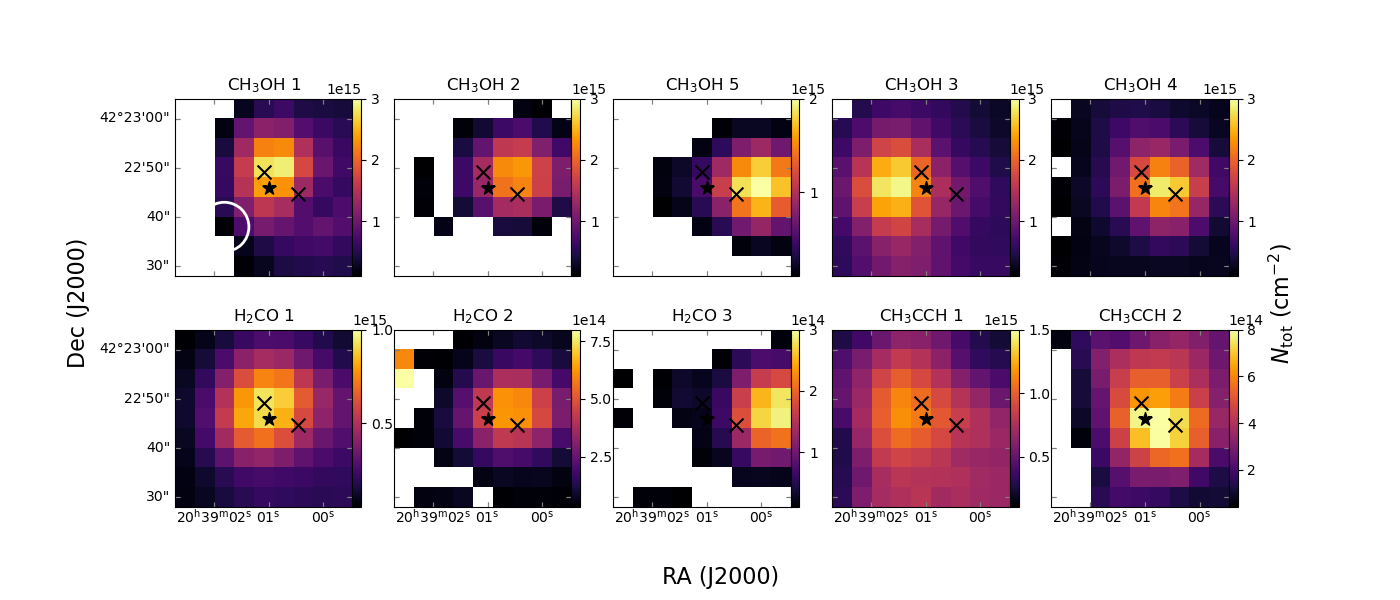}
\caption{Column density maps for all species (rows) in components (columns) of N44. The cores SMA 7 in MM1 and SMA 3 in MM2 are marked by x-es, and the clump of N44 is marked by a star (as designated in \cite{zapata2012} and \cite{motte2007}). CH$_3$OH 5 is moved to align with H$_2$CO 3, the broad components.}
\label{fig:n44ntot}
\end{figure*}

All three species required multiple components near DR21(OH)M, as broad and multi-peaked spectra are seen (Fig.~\ref{fig:n44ntot}).

We started with two components, N44 1 and N44 2, based on the $v_{\mathrm{lsr}}$ values of the dusty condensations MM1 (about $-$4.1 km s$^{-1}$) and MM2 ($-$0.7 km s$^{-1}$) provided by \citet{mangum1992}, \cite{mookerjea2012}, and \cite{minh2012}. In all three species we found components with similar characteristics, shown in Table~\ref{tab:ltemodel} and Fig.~\ref{fig:n44ntot}. Fig.~\ref{fig:n44lines} shows the spectra of the central pixel (marked by the black star in Fig.~\ref{fig:intintensity}), which is located in between MM1 and MM2 and shows multiple velocity components.

We find two cooler and extended components in CH$_3$OH 3 and 4 and CH$_3$CCH 1 and 2. We limit the region in Fig.~\ref{fig:n44ntot} to 18$^{\prime\prime}$, to encompass the MM1 and MM2 regions where the lines are strong enough to fit well. CH$_3$OH 3 (22~K) and CH$_3$CCH 1 (42~K) peak just southeast of MM1 and extend to most of the modelled N44 region. CH$_3$OH 4 (15~K) and CH$_3$CCH 2 (38~K) peak in between MM1 and MM2 and are slightly less extended. They all have the same $\Delta v = $ 2.8 km s$^{-1}$. In the interferometric observations of \cite{minh2012} and single dish observations of \cite{WidicusWeaver2017} there are CH$_3$OH components of 18-20~K. \cite{WidicusWeaver2017} also find one component of CH$_3$CCH at 37~K. At these temperatures, our cooler components likely trace the extended envelope around the MM1 and MM2 dusty condensations.

We find two warmer and more concentrated components in CH$_3$OH 1 and 2 and H$_2$CO 1 and 2. CH$_3$OH 1 (82~K) and H$_2$CO 1 (81~K) peak on MM1. CH$_3$OH 2 (75~K) and H$_2$CO 2 (84~K) are the most compact and are concentrated just north of MM2, showing only a few pixels of prominence. These components are broader than the cool components, with a $\Delta v = $ 4.0 km s$^{-1}$. Comparably, in H$_2$CO, \cite{zhao2024} find a kinetic temperature of 103~K for N44. Their spectral line profile clearly shows two velocity components (visually, similar to ours) but in their model they could not distinguish different components. In CH$_3$OH, \cite{minh2012} find a 200~K core component and \cite{WidicusWeaver2017} find a 93~K component. This suggests that at our spatial scales, on clump scales, we are likely smoothing out the hot gas near the SMA cores, thus the lower temperatures.

For the velocities of each of these components, we find a distinction between those peaking near MM1 or near MM2. Near MM1 the warm component is near -3.6 km s$^{-1}$ in CH$_3$OH and -4.1 km s$^{-1}$ in H$_2$CO, while the cold component is around $-$4.1-$-$4.2 km s$^{-1}$. Near MM2, the warm component is at $-$0.5 km s$^{-1}$ while the cold component is at $-$0.9 km s$^{-1}$. While our $v_{\mathrm{lsr}}$ values (Table~\ref{tab:ltemodel}) do not exactly match previous results, which contain components varying from $-$5 km s$^{-1}$ to 1 km s$^{-1}$ (\citealt{mangum1992,minh2012,girart2013,mookerjea2012}), the spatial scales over which we are modelling may provide slightly different average velocities. 


A broad feature that is blue-shifted relative to the source $v_\mathrm{lsr}$ of $-$4.1 km s$^{-1}$ is seen in both CH$_3$OH 5 and H$_2$CO 3. Fig.~\ref{fig:n44lines2} shows the spectra of a pixel about 6$^{\prime\prime}$ west of MM1 and MM2, where the outflow is most prominent. We fix this component at a $\Delta v $ of 15 km s$^{-1}$ and a $v_{\mathrm{lsr}}$ of $-$8 km s$^{-1}$. We are not able to reconcile the temperature of these two species in this component. H$_2$CO is fit well at 80~K, while CH$_3$OH is at 25~K. 

We first introduced this broad component based on the H$_2$CO and CH$_3$OH maps of \cite{zapata2012} Fig. 2. Their broad lines are emanating from SMA 6 and 7 with the blue- to red-shifted features going along an E-W gradient, not corresponding to our blue-shifted component. However, they also detect outflow emission in CO(2-1) and SiO, two classic outflow tracers, that do not correspond to their methanol outflow. The CO(2-1) emission features a high-velocity bipolar flow with its blue-shifted wing towards the west, and SiO shows a low-velocity flow with the same spatial pattern. This blue-shifted feature is also seen in CO and SiO in \cite{lai2003}, \cite{motte2007}, and \citet{schneider2010}. Our component, then, is likely tracing the CO outflow and is not associated with the maser outflow seen in \cite{vallee2006} or \cite{araya2009}. They see a red-shifted wing of a low-velocity outflow toward the NW which corresponds well to the maser observations of \cite{plambeck1990} and \cite{kogan1998}, the former of who suggest that the maser emission could be powered by outflows interacting with dense clumps of gas.


In the model spectrum (Figs.~\ref{fig:n44lines} and \ref{fig:n44lines2}), certain transition lines are underfit in CH$_3$OH. There is known non-thermal emission in this region, identified mainly through the prominent maser emission (\citealt{plambeck1990, kogan1998, kurtz2004, araya2009}). \cite{WidicusWeaver2017} found two transition lines at 230 and 250 GHz, which are not known as masers, with non-LTE behaviour and suggest they could be powered by similar mechanisms. Our lines at 290 GHz are similarly above the frequency at which any maser lines have been distinguished, but indicate non-LTE behaviour in our models.

\subsubsection{N44 RADEX modelling}

\begin{figure*}[h!]
\centering
\includegraphics[width=0.8\textwidth]{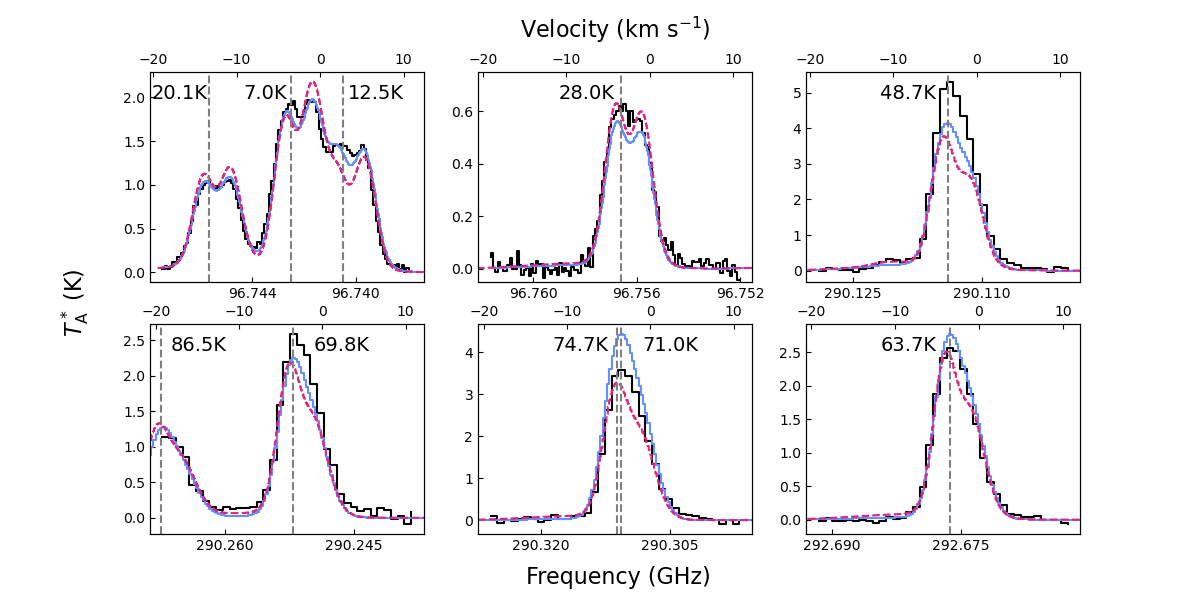}
\caption{The RADEX model spectrum for certain lines of CH$_3$OH in the centre of N44, in between MM1 and MM2. The data is in black, the RADEX model in solid blue, and the LTE model in dashed red. The upper energy level of each transition line is noted on the plot, with its frequency at the source $v_{\mathrm{lsr}}$ shown as the vertical grey line.}
\label{fig:n44radex}
\end{figure*}

\begin{figure*}[h!]
\centering
\includegraphics[width=0.8\textwidth]{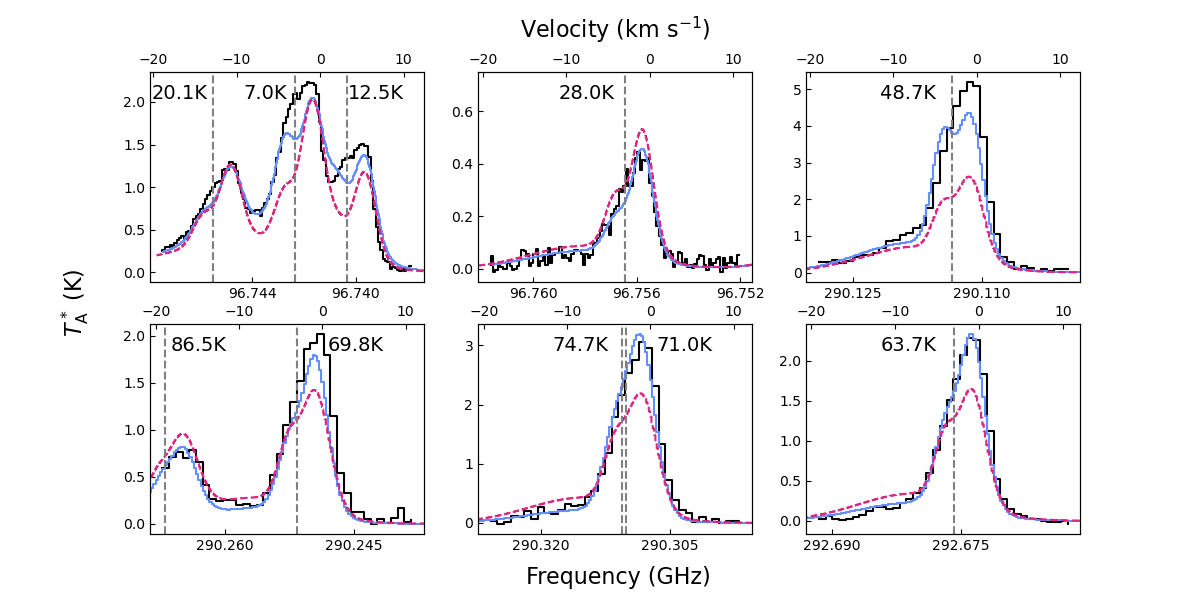}
\caption{Same as Fig.~\ref{fig:n44radex} for the outflow. This spectrum is where the outflow is most prominent, 6$^{\prime\prime}$ west from Fig.~\ref{fig:n44radex}.}
\label{fig:n44radex2}
\end{figure*}

To investigate the possible non-LTE behaviour of CH$_3$OH, we used the RADEX \citep{vandertak2007} module in CASSIS. Unlike the LTE model, RADEX is too computationally intensive to run over all pixels in the map. We chose to model the central pixel of the map as representative of MM1 and MM2 (Fig.~\ref{fig:n44radex}) and a pixel 6$^{\prime\prime}$ (three pixels) west of the central pixel in N44 where the outflow component is seen to be the most prominent (Fig.~\ref{fig:n44radex2}).

We used the same parameters as in the LTE model for the size, line widths, and velocities. Otherwise, the remaining fitted parameters are in Table~\ref{tab:radex}. We used the VASTEL Database\footnote{https://cassis.irap.omp.eu/?page=catalogs-vastel}, which uses the A- and E-CH$_3$OH forms, as there are only collisional coefficients for the separated A- and E- forms. These use p-H$_2$ as the collision species. The isotopic ratio was set at 1.

\begin{table*}
\caption{RADEX model results for CH$_3$OH in N44. The non-LTE $N_{\mathrm{tot}}$ represents that of A-CH$_3$OH only. Values without errors were fixed.}
\label{tab:radex}
\centering
\begin{tabular}{ l | l | c c c c  }
\hline\hline
Pixel & Component & LTE $N_{\mathrm{tot}}$ & $N_{\mathrm{tot}}$ & $T_{\mathrm{kin}}$ & $n_{\mathrm{p-H}_2}$ \\
& & (cm$^{-2}$) & (cm$^{-2}$) & (K) & (cm$^{-3}$)  \\
\hline
Centre & N44 1 & 2.38 ($\pm$0.13) $\times$ 10$^{15}$ & 1.67 ($\pm$0.11) $\times$ 10$^{15}$ & 78.0 & 6.0 $\times$ 10$^7$ \\
 & N44 2 & 1.30 ($\pm$0.09) $\times$ 10$^{15}$ & 7.76 ($\pm$0.24) $\times$ 10$^{14}$ & 100.0 & 5.0 $\times$ 10$^7$ \\
  & N44 3 & 2.21 ($\pm$0.07) $\times$ 10$^{15}$ & 7.23 ($\pm$0.43) $\times$ 10$^{14}$   & 22.8 ($\pm$4.1)  & 3.6 ($\pm$0.4) $\times$ 10$^7$ \\
    & N44 4 & 2.08 ($\pm$0.06) $\times$ 10$^{15}$ & 7.84 ($\pm$0.23) $\times$ 10$^{14}$ & 15.0 ($\pm$0.9)  & 2.9 ($\pm$0.6) $\times$ 10$^7$ \\
  & N44 5 & 5.22 ($\pm$0.45) $\times$ 10$^{14}$ & 2.34 ($\pm$0.36) $\times$ 10$^{14}$  & 19.4 ($\pm$1.7) & 6 $\times$ 10$^6$ \\
\hline
Outflow & N44 1 & 7.44 ($\pm$1.7) $\times$ 10$^{14}$ & 6.78 ($\pm$0.35) $\times$ 10$^{14}$ & 78.0 & 6.0 $\times$ 10$^7$ \\
 & N44 2 & 1.96 ($\pm$0.17) $\times$ 10$^{15}$ & 7.90 ($\pm$0.26) $\times$ 10$^{14}$ & 100.0 & 5.0 $\times$ 10$^7$ \\
  & N44 3 & 7.70 ($\pm$0.67) $\times$ 10$^{14}$ & 3.96 ($\pm$0.20) $\times$ 10$^{14}$ & 35.0 ($\pm$0.8) & 1.5 ($\pm$0.3) $\times$ 10$^6$ \\
    & N44 4 & 1.65 ($\pm$0.08) $\times$ 10$^{15}$ & 8.05 ($\pm$0.55) $\times$ 10$^{14}$ &  24.4 ($\pm$0.9) & 1.2 ($\pm$0.2) $\times$ 10$^7$ \\
  & N44 5 & 2.13 ($\pm$0.09) $\times$ 10$^{15}$ & 9.53 ($\pm$0.20) $\times$ 10$^{14}$ & 23.8 ($\pm$0.5) & 6 $\times$ 10$^6$ \\

\hline
\end{tabular}
\end{table*}

The results show CH$_3$OH is near LTE with most excitation and kinetic temperatures within 1-10~K of each other. In the centre pixel (Fig.~\ref{fig:n44radex}) most lines are just as well represented by the LTE model. In the outflow (Fig.~\ref{fig:n44radex2}), certain lines are represented much better in the RADEX model. There, the GBT triplet at 96.74 GHz (E$_{\mathrm{up}}$ = 20, 7, 13~K), and the IRAM lines at 290.1, 290.3, and 292.7 GHz (E$_{\mathrm{up}}$ = 49, 75, 71, 64~K). The GBT lines, which are dominated by the colder components in LTE, are likely tracing sub-thermally excited gas. These GBT and IRAM lines show visually the most prominent outflow wings, and could have non-thermal excitation due to the shock conditions.

We tried running the LTE model with adjusted temperatures from these RADEX results. In component 2, a higher temperature does not make a clear difference. In components 3 and 4, a higher temperature from the outflow results produces worse results. Similarly, an adjusted temperature in component 5 produces worse results. However, components 3 and 4 peak near the centre pixel, where the LTE and RADEX temperatures align, and component 5 peaks near the outflow pixel, where similarly the LTE and RADEX temperatures align.

\subsection{N48}

\begin{figure}[h!]
\centering
\includegraphics[width=0.5\textwidth]{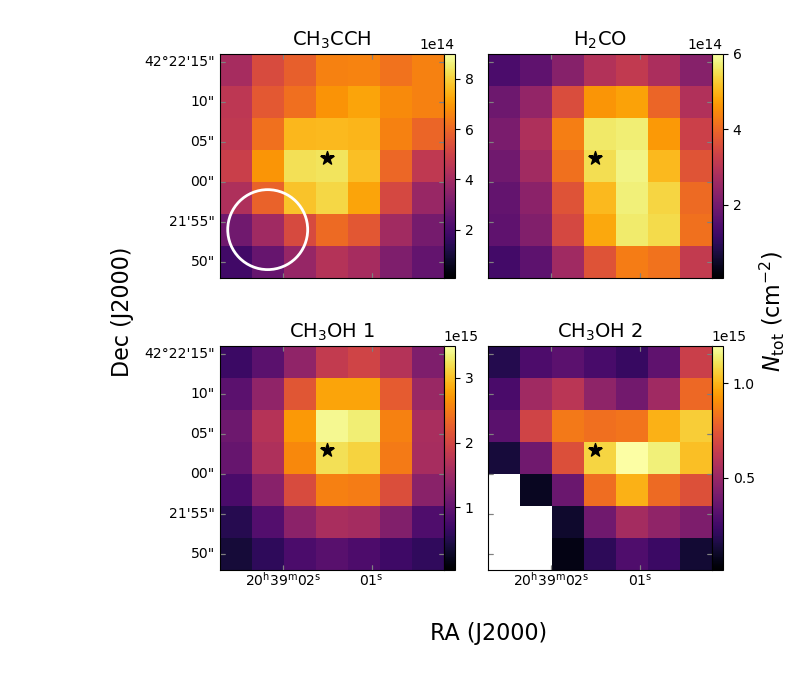}
\caption{Column density maps for all species in components of the region N48, where the clump location is indicated by a star (\citealt{motte2007}).}
\label{fig:n48ntot}
\end{figure}

N48 is a well-studied clump, with \cite{mangum1991} showing it also as a high luminosity source with a potentially embedded star and \cite{motte2007} classifying N48 as a massive IR-quiet protostellar source.
In our results, for CH$_3$CCH and H$_2$CO a one-component fit worked well for most of the map (Fig.~\ref{fig:n48ntot}). These were found to be 34~K and 68~K, respectively. 

In CH$_3$OH, we identified two components, with one at 20~K and the other 8~K (Table~\ref{tab:ltemodel}). They had equal line widths of 3.2 km s$^{-1}$ and we fixed the velocity of the second component to that of the first component. The region is known to have structure on smaller scales -- with the Plateau de Bure Interferometer \cite{bontemps2010} see 5 fragments in 1 mm emission and 2 main fragments in 3 mm emission, while \cite{csengeri2011b} find two components in H$^{13}$CN at a $v_{\mathrm{lsr}}$ of $-$2.39 and $-$4.77 km s$^{-1}$. Our two components are at $-$4.04 and $-$3.78 km s$^{-1}$, similar, but again likely offset due to the spatial scales and environments we can trace.

\cite{mangum1991} see an E-W extension in the source. Our second component of CH$_3$OH (Fig.~\ref{fig:n48ntot}) has a westward extension, and towards the edge of the modelled region moves towards the northwest, towards N38.

\subsection{N38}

\begin{figure}[h!]
\centering
\includegraphics[ width=0.5\textwidth]{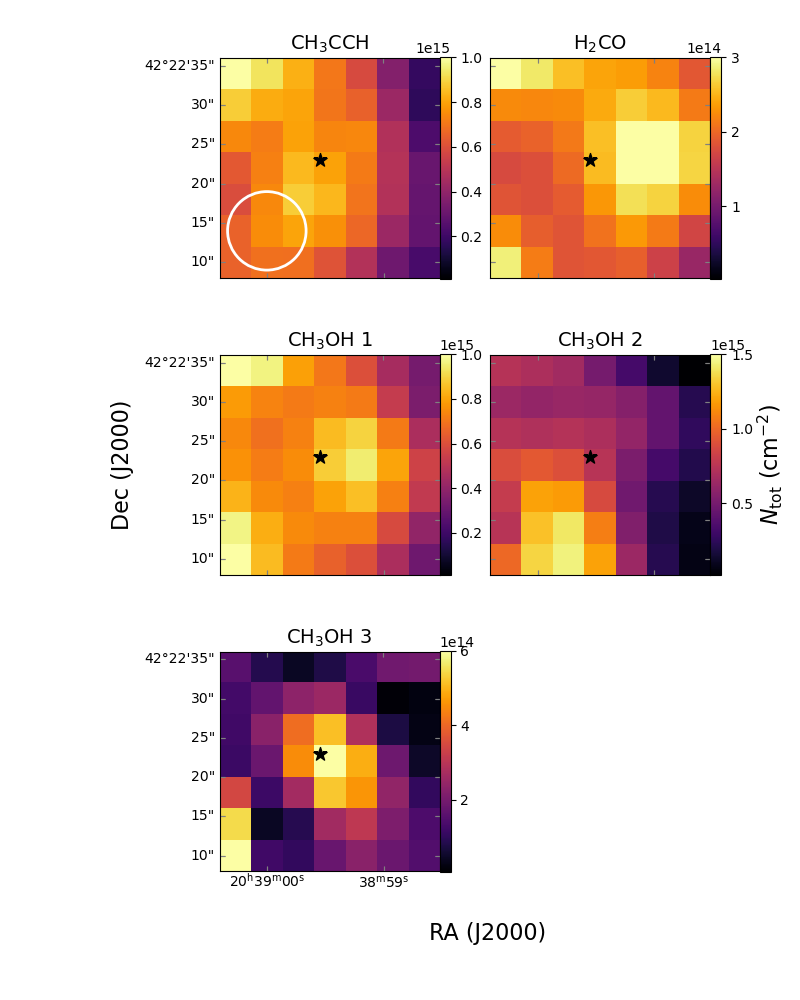}
\caption{Same as Fig.~\ref{fig:n48ntot} for the clump N38.}
\label{fig:n38ntot}
\end{figure}

N38 is seen as a luminous source with a potentially embedded star and a N-S extension, according to \cite{mangum1991}, with \cite{motte2007} classifying it as a massive IR-quiet protostellar source. In our data, we identify one component each in CH$_3$CCH and H$_2$CO and three components in CH$_3$OH (Figs.~\ref{fig:n38ntot} and~\ref{fig:n38lines}).

Similar to the previous clumps, we are seeing a warm, extended clump. H$_2$CO peaks at 53~K just west of the \cite{motte2007} clump peak while the CH$_3$CCH component peaks just south of the clump at 29~K. This is similar to other temperature derivations of the region, where \cite{mangum1992} find two components at 26 and 25~K in NH$_3$ and \cite{zhao2024} find a H$_2$CO kinetic temperature of 57~K. We are not necessarily smoothing out a hot core here.

In CH$_3$OH, we see one warm, 20~K peak that dominates the fit in the IRAM lines and spatially matching the H$_2$CO peak (Fig.~\ref{fig:n48ntot}). We also find two cold, 8~K, components that dominate the fit in the GBT lines. One of these peaks on the clump and one to the southeast. This southeast clump aligns spatially with the cold CH$_3$OH component that extends from N48 to the northwest (Fig.~\ref{fig:n48ntot}). We infer this cold gas is coming from the extended filament of the DR21 ridge. Similarly, \cite{csengeri2011} find this cold, extended gas in between N38 and N48 with single dish observations of N$_2$H$^+$. The cold gas we observe peaking on the clumps of N38 and N48 are only seen by \cite{csengeri2011} with interferometric observations of N$_2$H$^+$.

\subsection{N - N40, N41, N36}

\begin{figure}[h!]
\centering
\includegraphics[width=0.5\textwidth]{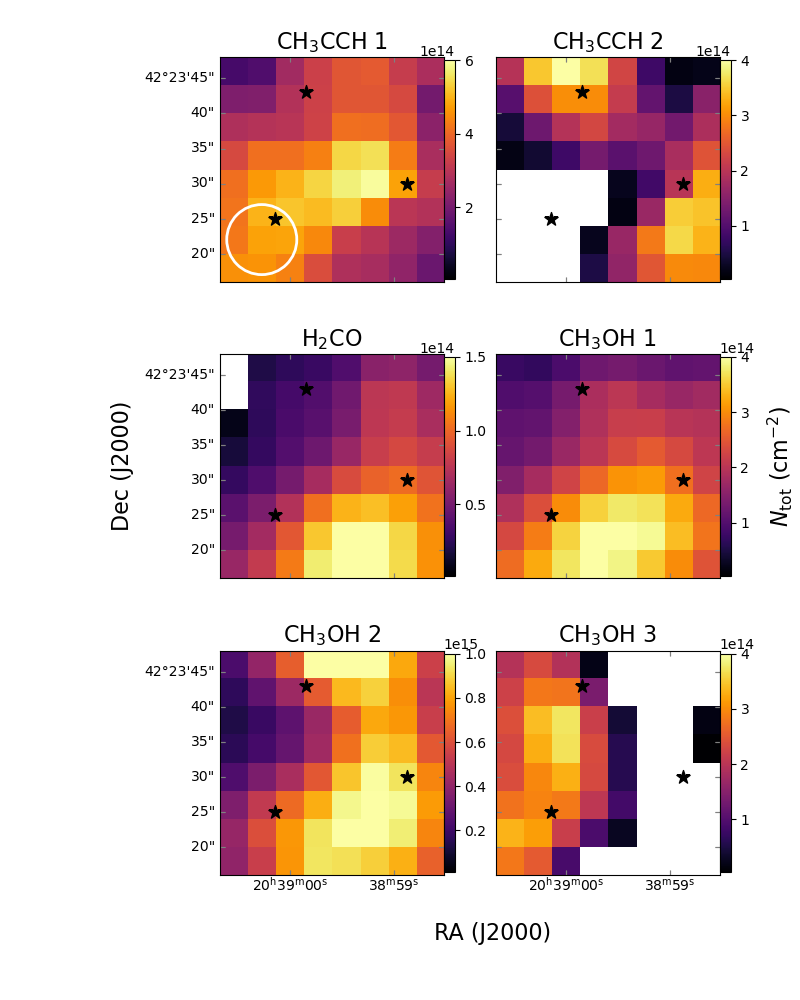}
\caption{Same as Fig.~\ref{fig:n48ntot} for the N region, containing the clumps N40, N41, and N36.}
\label{fig:n36ntot}
\end{figure}

Towards the north region N, we grouped together N40 (Fig.~\ref{fig:n40lines}), N36 (Fig.~\ref{fig:n36lines}), and N41 (Fig.~\ref{fig:n41lines}). The column density maps are in Fig.~\ref{fig:n36ntot}, where we find H$_2$CO fit well by one component, while CH$_3$CCH required two components: one 30~K component with a line width of 2.6 km s$^{-1}$, central in between the three clumps identified by \cite{motte2007}, and a 25~K component with a line width of 1.2 km s$^{-1}$ towards the outer regions of N41 and N36. In CH$_3$OH, we identified three components: one warm, 27~K component and two cold, 6~K components, with one moderately broad at 3.2 km s$^{-1}$ and one narrow at 1.2 km s$^{-1}$ (see also Table~\ref{tab:ltemodel}). The velocities were fixed at that of the first component.

\cite{motte2007} classify N40 as a massive IR-quiet protostellar source from 1.2 mm continuum emission, yet \cite{bontemps2010} note it is weak at 3 mm and may be extended, unlike the other dense clumps they study. In comparison to our other clumps, we also see weaker emission in this region (Fig.~\ref{fig:intintensity}).

\section{Chemical modelling - results and analysis}
\label{section:discussion}

The components distinguished by our LTE model show the large-scale environment around many, known core-scale features. DR21(OH), within the larger DR21 ridge, is a young complex with embedded protostars that have not yet disrupted the parent cloud.

In order to determine the possible origin of these molecules in the observed environments, we modelled the species abundances using the NAUTILUS gas-grain code \citep{ruaud2016}. We did this for all species in all components of DR21(OH), or N44 (we do not use the N44 designation for each component in this section, as all are N44). NAUTILUS uses three phases for chemical evolution -- gas, grain surface, and grain mantle -- and stimulates the abundances over time as a function of different physical conditions -- gas density, dust and gas temperature, visual extinction, ultraviolet flux and cosmic ray ionisation. In the code, coupled differential equations describe a network of over 10,000 reactions connecting 489 species. 

Before starting the model, we found the fractional abundances, or the ratio of the column density for a given species and the column density of molecular hydrogen, H$_2$. For our species, we used the modelled column density (Sect.~\ref{sect:n44}) at its peak value. We found the H$_2$ column density at the same position by re-gridding and smoothing the \cite{cao2022} H$_2$ map in CASA\footnote{https://casa.nrao.edu/} onto our grid and beam size. \cite{cao2022} derived their map using archival data described in \cite{cao2019}: Herschel/PACS 160 $\mu$m and James Clerk Maxwell Telescope (JCMT)/SCUBA-2 450 and 850 $\mu$m images.

Studies of the H$_2$ to $A_V$ conversion range from $A_V$ = $\frac{N_{H_2}}{2.1\times10^{21}}$ (\citealt{rachford2009, zhu2017}) to $A_V$ = $\frac{N_{H_2}}{1.9\times10^{21}}$ (\citealt{bohlin1978, whittet1981}). In this paper, we assumed an `average' value of $A_V$ = $\frac{N_{H_2}}{2\times10^{21}}$. We used a standard cosmic ray ionisation rate of $1.3\times10^{-17}$ s$^{-1}$. For the UV field, the NAUTILUS code provided a function of the form $S\times10^8$~photons~cm$^{-2}$~s$^{-1}$ \citep{ruaud2016} where S is a scaling factor. We used a scaling factor of 1 to model a general interstellar radiation field. We note that this could be higher in the DR21 environment, however, given the large H$_2$ column densities and resulting values of $A_V$, larger UV field strengths are unlikely to have any effect on the chemistry due to shielding.

We used the volume densities, $n_{\mathrm{p-H_2}}$, similar to the CH$_3$OH RADEX models. We matched the other species components as: CH$_3$OH components 1 and 2 at 1 $\times$ 10$^{8}$ to H$_2$CO components 1 and 2; CH$_3$OH Component 5 at 5 $\times$ 10$^6$ to H$_2$CO component 3; CH$_3$OH components 3 and 4 at 2 $\times$ 10$^7$ to CH$_3$CCH components 1 and 2. This comparison is based on the temperature, line width, and velocity parameters found for each species (Table~\ref{tab:ltemodel}). Such high densities are not unusual for a dense core (e.g. \citealt{mookerjea2012, girart2013}) and are representative of (near) LTE conditions.


We started with a two-phase model in NAUTILUS: a cold quiescent cloud followed by a warm-up phase powered by embedded protostars. The cold cloud had initial abundances representing the diffuse ISM, a temperature of 10~K, a density of $10^4$~cm$^{-3}$, a visual extinction of 50 magnitudes, and it evolved for 10$^5$ years (see \citealt{ruaud2016}). Then, in the warm-up phase the gas and dust temperature is the observed temperature of each pixel found from the LTE model, and the visual extinction is calculated from the average H$_2$ column density. This evolved for another 10$^6$ years.

\begin{figure*}[h!]
\centering
\includegraphics[trim = 1cm 0 1cm 0, clip=True, scale=0.5]{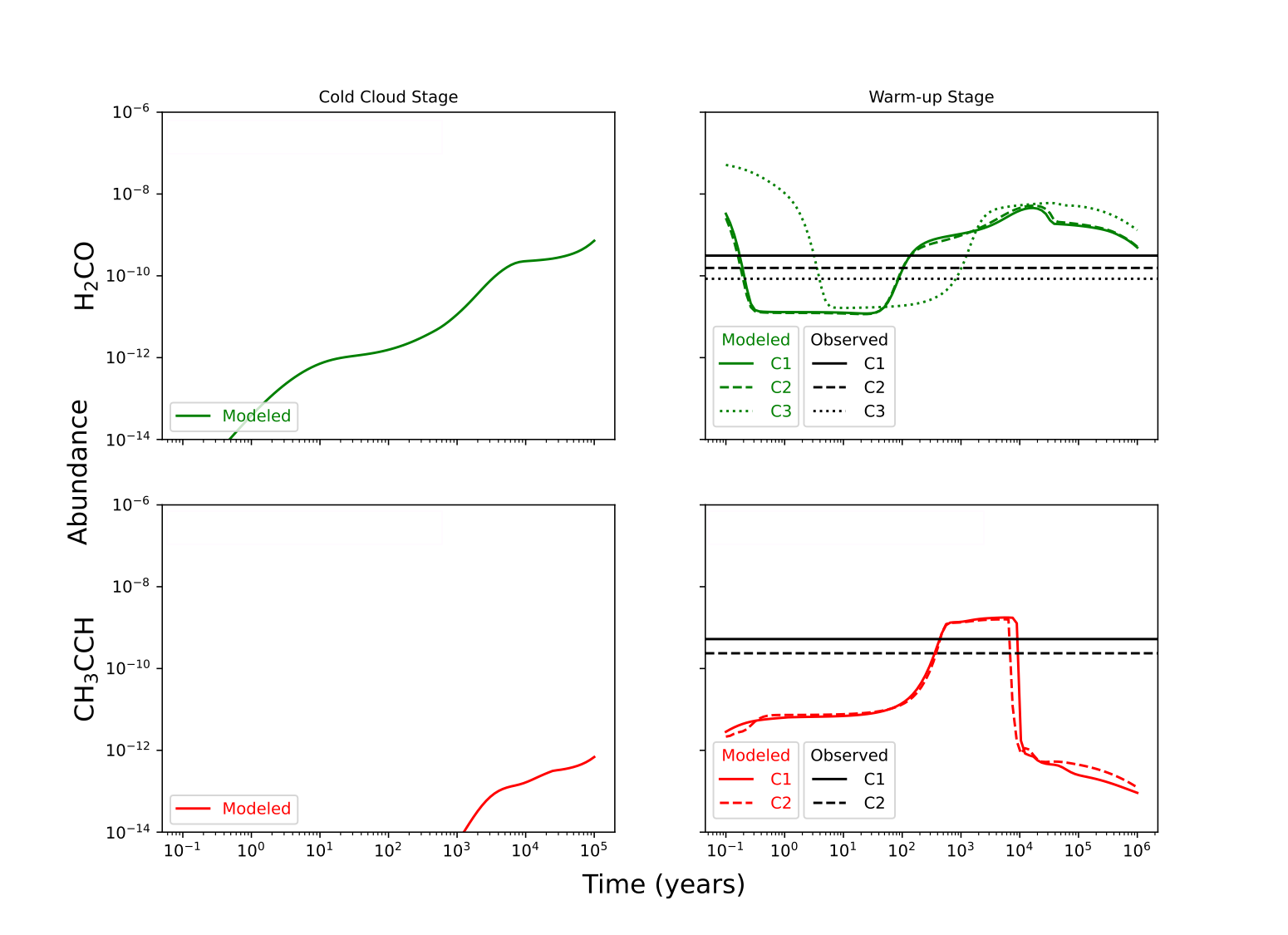}
\caption{The two-stage model for H$_2$CO (top) and CH$_3$CCH (bottom). The 10~K cold cloud stage (left) is followed by a warm up stage where the LTE components are separated based on the temperatures and densities observed. The time axis represents the age of each component independently. The components of each model are in dotted, dashed, or solid lines, where the coloured lines are the model abundances and the black lines are the observed abundances calculated from the averages in Table~\ref{tab:ltemodel}.}
\label{fig:2stage}
\end{figure*}

For all three components of H$_2$CO, the warm-up phase was adequate to reproduce our observed gas-phase abundances (Fig.~\ref{fig:2stage}, top). For components 1 and 2, it takes just over 100 years to produce the observed abundances, while for component 3, the assumed outflow, it takes over 10$^3$ years for the abundances to match. The same chemistry applies to all components, on different timescales -- at the time the abundances match, the domination production mechanism changes (Figs.~\ref{fig:proddesth2co1}, \ref{fig:proddesth2co3}). The main production mechanisms at the start are the gas phase reactions:
\begin{equation}
    \mathrm{O} + \mathrm{CH_3} \rightarrow \mathrm{H} + \mathrm{H_2CO},
    \label{eq:h2co}
\end{equation}
and
\begin{equation}
    \mathrm{CH} + \mathrm{H_2O} \rightarrow \mathrm{H} + \mathrm{H_2CO}.
\end{equation}
Then, these are replaced by desorption of H$_2$CO off the grain surfaces. For all components, the destruction mechanism follows a similar timeline. H$_2$CO is, at first, solely destroyed by the gas phase reaction:
\begin{equation}
    \mathrm{H_2CO} + \mathrm{C} \rightarrow \mathrm{CH_2} + \mathrm{CO},
\end{equation}
and after 100 or 10$^3$ years, depending on the component, the dominant destruction mechanism becomes adsorption onto the grain surface. At this point, the abundance of gas phase C has sharply decreased.

Overall, the temperatures of $\sim$80~K are enough to produce the observed abundances. The binding energy of H$_2$CO is reported as 3260~K, which corresponds to a desorption temperature of 110~K \citep{Penteado2017}. At smaller scales, \cite{zhao2024} report the H$_2$CO kinetic temperature to be 103~K, so we are likely seeing the hot core smoothed out, and diluting the temperature with the cooler, extended gas.

As CH$_3$CCH is seen as an envelope molecule, we tried simulating a cloud-edge environment with an $A_V$ of 1 (Fig.~\ref{fig:2stage}, bottom). In this scenario, the chemical evolution is influenced by the external interstellar radiation field (e.g. \citealt{spezzano2016, Kalvans2021}). The abundances for both components are reached at a few hundreds of years after the cold cloud stage. At 10$^4$ years, there is a sharp decrease in abundance.

The main production route (Fig.~\ref{fig:proddestch3cch}), from about 20 years on, is the grain surface hydrogenation:
\begin{equation}
    \mathrm{gr(H)} + \mathrm{gr(CH_2CCH)} \rightarrow \mathrm{CH_3CCH}.
\end{equation}
Prior, it's the grain surface reaction:
\begin{equation}
    \mathrm{gr(CH)} + \mathrm{gr(C_2H_3)} \rightarrow \mathrm{CH_3CCH}.
\end{equation}
Relatively, the grain surface production is largely dominant over gas phase production (as in \citealt{hickson2016, ines2018, calcutt2019}). This is not consistent with CH$_3$CCH being formed in a WCCC scenario. Once the abundance peaks, the main destruction route is photodissociation to H and CH$_2$CCH -- assumed to be influenced by the low $A_V$ environment.

However, DR21(OH) is located within a clustered region, and we do not know if the extent of CH$_3$CCH is truly on the envelope edge. Values of $A_V$ up to $\sim$5 still worked in our model if the UV field in DR21 is larger than unity. However, modelling CH$_3$CCH with a higher $A_V$ corresponding to cloud core conditions (as in H$_2$CO, $A_V$=50) failed to reproduce the observed abundances at any value of the UV field. In this case, a shock -- a sharp spike in temperature and density -- was needed to reproduce the CH$_3$CCH abundances. This model was also introduced for CH$_3$OH, and is described next.

For CH$_3$OH, the warm-up model, with any value of $A_V$ or $n_{\mathrm{p-H_2}}$, did not reproduce the abundances. We then used the approach as in \cite{freeman2023} and simulated a shock passing through the cloud. Shocks, which can be produced by outflows such as the ones in our source, can liberate grain species into the gas phase causing a sharp increase in gas phase abundances (\citealt{herbst2009, palau2017}). While we cannot model a shock directly, we simulate shock-induced, non-thermal desorption by implementing a three stage model: a cold quiescent cloud, a `shock' with a sharp increase in temperature and density, and a post-shock phase where the gas settles to the observed conditions.

\begin{figure*}[h!]
\centering
\includegraphics[trim = 1cm 0.1cm 1cm 0, clip=True, scale=0.5]{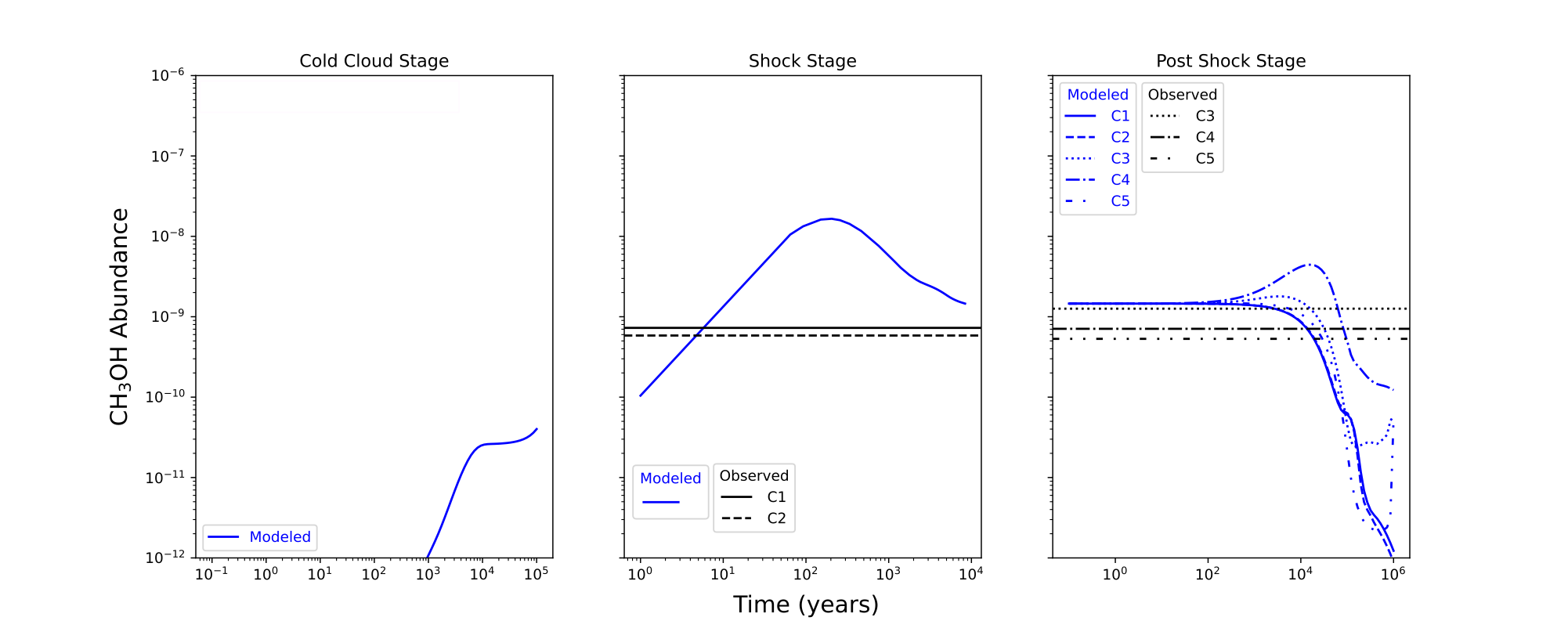}
\caption{The three-stage model for CH$_3$OH, simulating a shock. The 10~K cold cloud stage (left) is followed by a sharp increase in density and temperature to simulate a shock (middle) and then a post-shock stage (right), where the gas settles to the LTE conditions observed. The time axis represents the age of each component independently. The LTE components are separated based on their temperatures and densities. The components of each model are in dotted, dashed, or solid lines, where the coloured lines are the model abundances and the black lines are the observed abundances calculated from the averages in Table~\ref{tab:ltemodel}. As C1 and C2 are warmer, we suggest they are still in the shock stage, whereas the cooler C3, C4, and C5 are in a post-shock state.}
\label{fig:shock}
\end{figure*}

The cold cloud is the same as the previous models. In the shock stage, the gas temperature and density evolve as in \citet{palau2017} for IRAS 20126 (see their Fig. 5, top). This assumes a C-type shock as in \cite{jimenezserra2008} with a shock velocity of 40~km s$^{-1}$ and a pre-shock density of $10^4$~cm$^{-3}$. This lasts for 10$^4$ years, during which the temperature sharply increased to 2500~K before decreasing, while the density slowly increased from $10^4$ to $8.2\times10^4$~cm$^{-3}$, where it stabilised. We fix the dust temperature to 80~K. Then, in the post-shock phase, we return the temperature to those observed from the LTE model (as used in the warm-up stage above), set the density at 10$^5$ cm$^{-2}$ and allowed the cloud to evolve for a final $10^6$ years. For all five components, this reproduced the abundances (Fig.~\ref{fig:shock}).

In component 2, the warm 80~K region, the main production route is the gas phase reaction (Fig.~\ref{fig:proddestch3oh2}):
\begin{equation}
    \mathrm{CH_3OCH_3} + \mathrm{C} \rightarrow  \mathrm{CH_3OH} + \mathrm{C_2H_2}.
\end{equation}
After 10$^5$ years the dominant mechanism sharply changes to:
\begin{equation}
    \mathrm{CH_3OCH_4^+} + \mathrm{e^-} \rightarrow \mathrm{CH_3} + \mathrm{CH_3OH}.
    \label{eq:dissrecomb}
\end{equation}
We assume component 1 is similar from their physical parameters (Table~\ref{tab:ltemodel}). Due to their higher temperature compared to the other components, we assume these are still in a shock stage in Fig.~\ref{fig:shock}, where their abundances match the model after less than 10 years.

For both the cold and narrow component 4 and the cold and broad component 5, grain surface production dominates through:
\begin{equation}
    \mathrm{gr(H)} + \mathrm{gr(CH_3O)} \rightarrow \mathrm{CH_3OH},
\end{equation}
and
\begin{equation}
    \mathrm{gr(H)} + \mathrm{gr(CH_2OH)} \rightarrow \mathrm{CH_3OH}.
\end{equation}
This is dominant over 10$^6$ years in component 4. For component 5, after 10$^4$ years, the gas phase dissociative recombination (Eq.~\ref{eq:dissrecomb}) dominates. In the model, we assume these colder components are in the post-shock stage in Fig.~\ref{fig:shock}. Their model abundances are slightly overproduced until about 10$^5$ years, where they sharply decrease. In the cold components, it is clear that some non-thermal desorption mechanism after grain surface production is key to producing the gas phase abundances. 

For all components, the main destruction route is adsorption to the grain surface until about 10$^5$ years. After that, in all components, there are contributions from gas phase ion-molecule reactions. 

To see if the timescales are physically realistic, we look at the outflow. For a flow moving at 7 km s$^{-1}$ (from component N44 5, Table~\ref{tab:ltemodel}), to span 10$^{\prime\prime}$ (Fig.~\ref{fig:n44ntot}, \citealt{zapata2012} Fig. 2), it would need to be 1.6 $\times$ 10$^4$ years old. This is less than the time it takes for all five components to match model and observed abundances in the post-shock stage. 

We found that in higher density models the same evolution proceeded on a faster timescale. We tried to use the RADEX densities in the shock model as we had for the warm-up models to reconcile the results of our different analyses. If the RADEX densities were used in the post-shock stage, the abundances would match at a few 10$^3$ years for components 3, 4, and 5. However, there would be a sharp density jump from the shock stage to the post-shock stage in this scenario, for which we have no explanation. We tried using the RADEX densities in the shock stage, and the CH$_3$OH was destroyed quickly. However, along with the abundances not matching, we have no basis for such high densities in the shock stage as we use a previously parameterised shock. Different type shocks have been explored for example in L1157 \citep{james2020}, but it is out of the scope of this paper. We can only conclude that CH$_3$OH need some form of non-thermal desorption to produce the observed abundances.

We can speculate on possible sources powering non-thermal desorption in DR21(OH). There are numerous outflows seen emanating from MM1 and MM2, including the previously discussed molecular blue-shifted outflow, and the E-W CH$_3$OH maser outflows which contain other species such as H$_2$CO and H$^{13}$CO$^+$ (\citealt{zapata2012, OrozcoAguilera2019}) as well as thermal CH$_3$OH emission. A N-S H$_2$CS outflow is seen from MM1 with rotation temperatures of 40-60~K \citep{minh2011}, where lines of CH$_3$OH are also detected \citep{minh2012}. They suggest CH$_3$OH could arise from the same mechanism as H$_2$CS: non-thermal desorption from outflows interacting with the surrounding gas. While we cannot discern the exact sources of the non-thermal desorption in the CH$_3$OH components, the clustered cores of MM1 and MM2 are evidently producing multiple sites of turbulent gas. Further modelling of higher spatial resolution data is needed to determine the structure of DR21(OH) and connect it to the potential molecular formation routes.


\section{Summary}
\label{section:summary}

We presented IRAM and GBT observations mapping DR21(OH) at 10$\arcsec$ resolution. We used CH$_3$CCH, CH$_3$OH and H$_2$CO emission lines to determine the physical parameters of the different gaseous environments in the region, connecting the core scales to the larger clump and filaments scales. Our main findings are that:
\begin{itemize}
    \item DR21(OH) main, or N44, is found to have a $\sim$80~K, and $\sim$30~K components near both MM1 and MM2, interpreted as the warmer inner core and the cooler envelope. H$_2$CO and CH$_3$OH also show a broad, blue-shifted outflow component that is not associated with the maser outflows of this region.
    \item chemical modelling indicates that H$_2$CO can be produced in the warm-up stage through thermal mechanisms, CH$_3$CCH can be produced in the warm-up stage in an environment with low $A_V$, and CH$_3$OH needs a non-thermal desorption mechanism such as a shock to produce the observed abundances. For CH$_3$CCH and CH$_3$OH especially, grain surface production is important to reproducing their gas phase abundances.
    \item other dusty condensations in the region show warm, $\sim$30~K, gas with similar line widths and velocities as that of N44. The cold components seen are likely non-LTE, and represent sub-thermally excited gas.
\end{itemize}
With a sophisticated LTE model we are able to use large-scale single-dish data to disentangle different temperature and velocity components of star-forming clumps and cores. This is useful even for complicated species such as CH$_3$OH, as we are able to apply the model over the whole map along with targeted use of the more computationally expensive RADEX. The LTE model allows us to confidently determine the abundances of each of these species, essential for determining the chemical evolution of the region.

\newpage

\begin{acknowledgements}

This work is based on observations carried out under project numbers 021-20 and 122-20 with the 30m telescope. IRAM is supported by INSU/CNRS (France), MPG (Germany) and IGN (Spain). The Green Bank Observatory is a facility of the National Science Foundation operated under cooperative agreement by Associated Universities, Inc.

PF would like to thank L. Morgan and D. Frayer at the GBT/NRAO for their help developing observing scripts as well as GBTIDL scripts for ARGUS calibration and reduction, as well as P. Torne and M. Rodriguez at IRAM for their help observing with the IRAM telescope. PF and RP would like to thank K. Qiu and Y. Cao for providing the H$_2$ columnn density maps.

PF and RP acknowledge the support of the Natural Sciences and Engineering Research Council of Canada (NSERC), through the Canada Graduate Scholarships – Doctoral, Michael Smith Foreign Study Supplement, and Discovery Grant programs.

As researchers at the University of Calgary, PF and RP acknowledge and pay tribute to the traditional territories of the peoples of Treaty 7, which include the Blackfoot Confederacy (comprised of the Siksika, the Piikani, and the Kainai First Nations), the Tsuut’ina First Nation, and the Stoney Nakoda (including Chiniki, Bearspaw, and Goodstoney First Nations). The City of Calgary is also home to the Métis Nation of Alberta Districts 5 and 6.

\end{acknowledgements}

\bibliographystyle{aa} 
\bibliography{bib}

\begin{appendix}
\onecolumn

\section{Model spectral line fits}

\begin{figure*}[h!]
\centering
\includegraphics[scale=0.5]{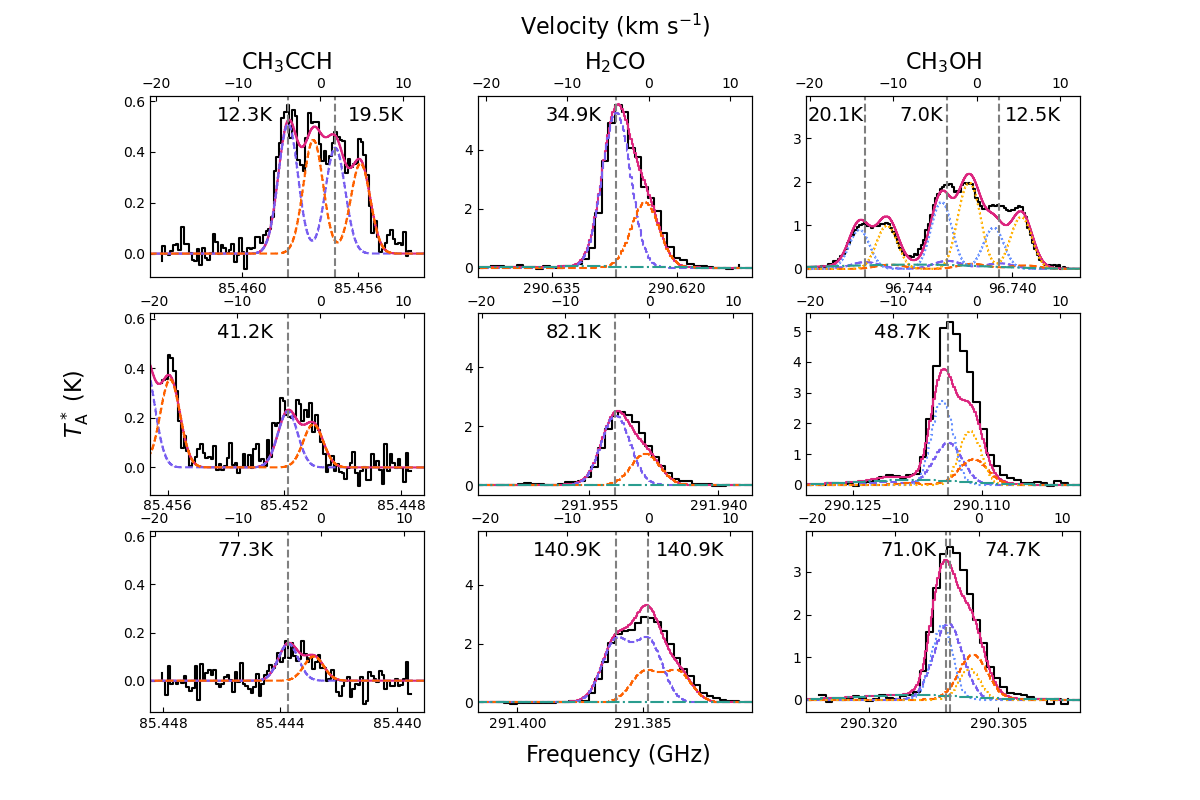}
\caption{The LTE model spectrum for each of the three species in the central pixel of N44. The data is in solid black, the overall model is in solid pink. The components are indicated by dashed or dotted coloured lines: the dashed orange and purple are the warm components of each species, the dotted yellow and blue are the cold components in CH$_3$OH, and the dot-dashed teal is the broad outflow component. The upper energy level of each transition line is noted on the plot, with its frequency at the source $v_{\mathrm{lsr}}$ shown as the vertical grey line.}
\label{fig:n44lines}
\end{figure*}

\begin{figure*}[h!]
\centering
\includegraphics[scale=0.5]{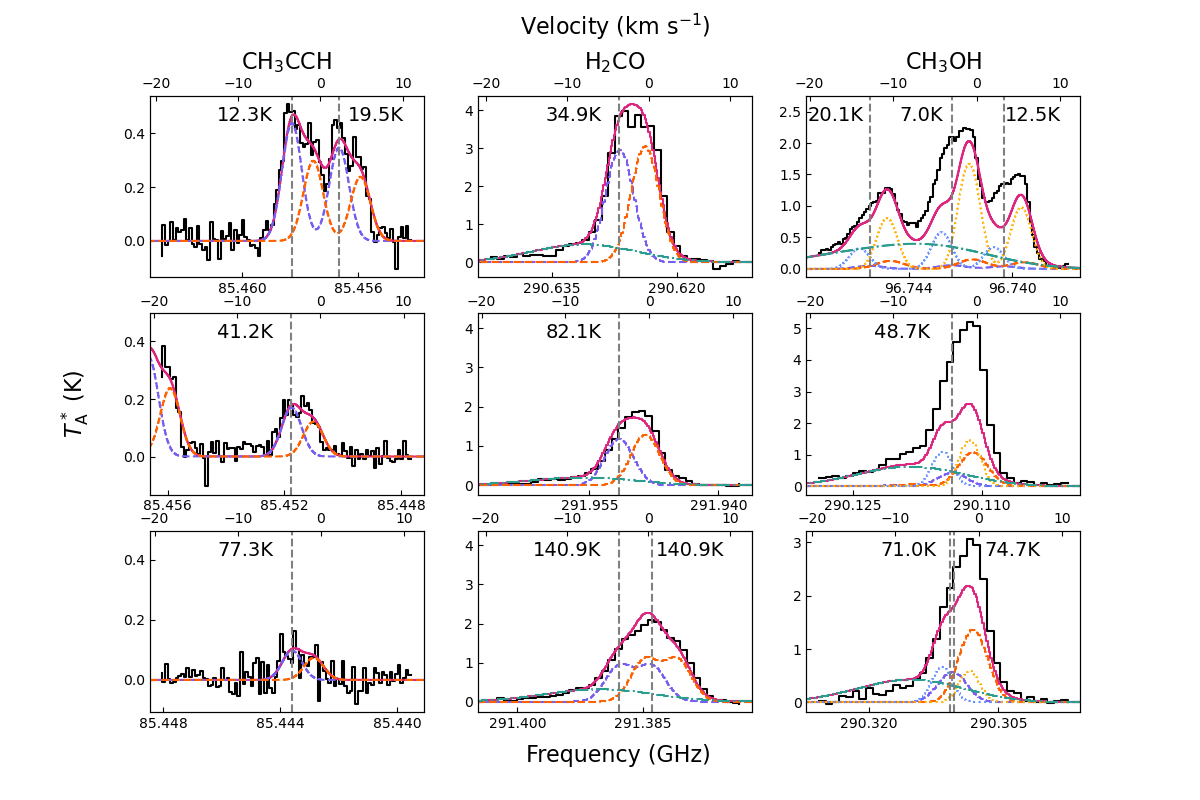}
\caption{Same as Fig.~\ref{fig:n44lines} for the outflow component of N44. This pixel is three pixels, or 6$^{\prime\prime}$, west of the central pixel in N44, and is where the outflow component is seen to be the most prominent.}
\label{fig:n44lines2}
\end{figure*}

\begin{figure*}[h!]
\centering
\includegraphics[scale=0.5]{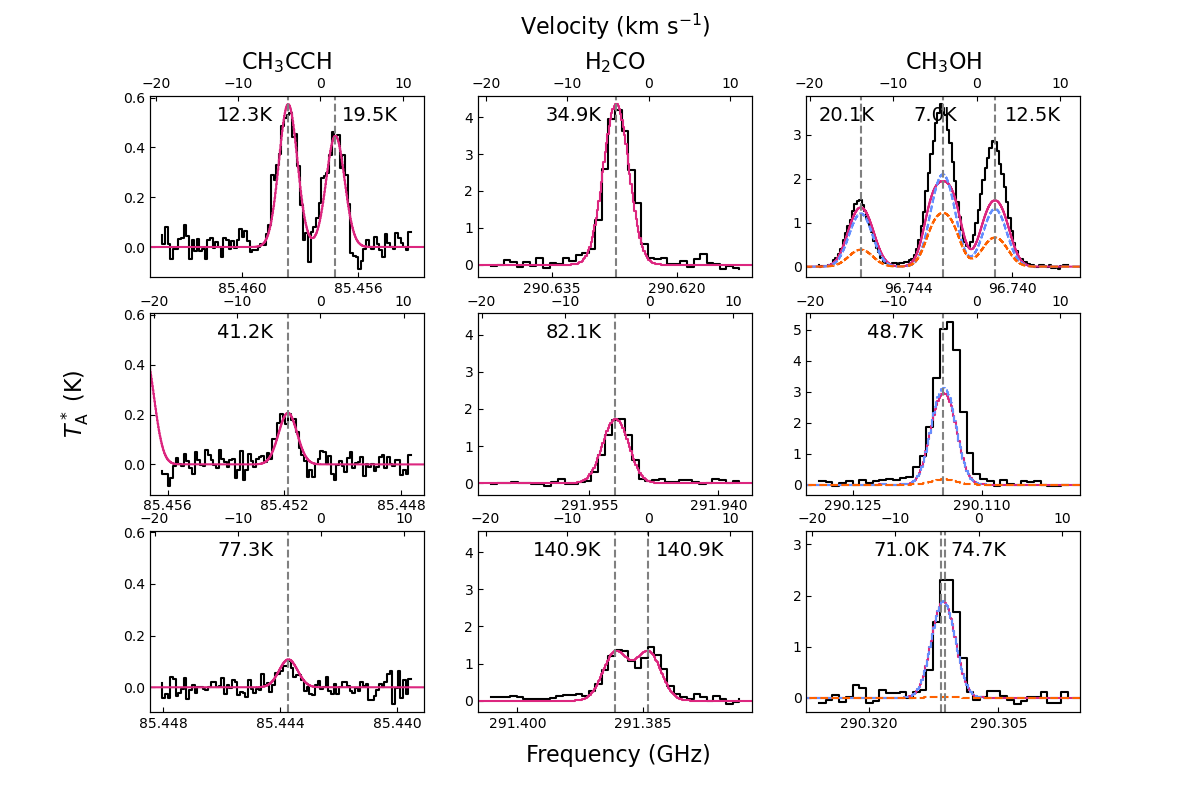}
\caption{Same as Fig.~\ref{fig:n44lines} for the central pixel of N48.}
\label{fig:n48lines}
\end{figure*}

\begin{figure*}[h!]
\centering
\includegraphics[scale=0.5]{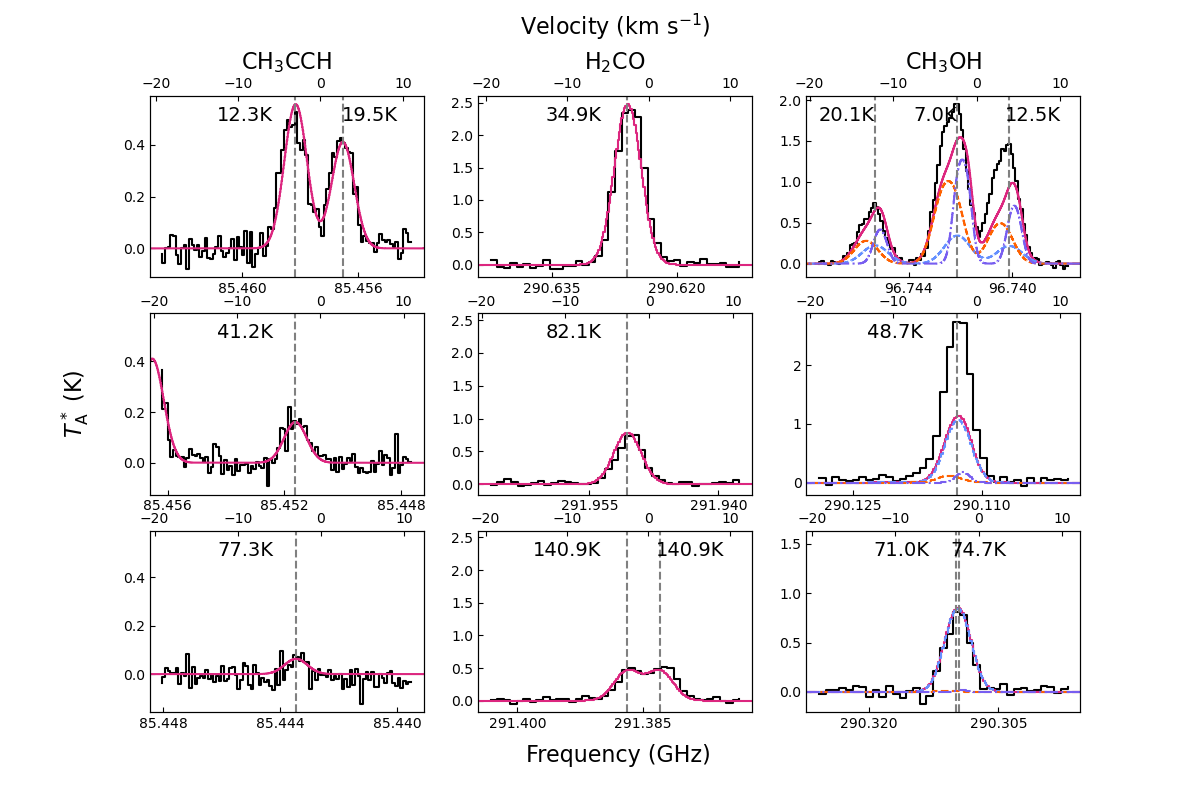}
\caption{Same as Fig.~\ref{fig:n44lines} for the central pixel of N38.}
\label{fig:n38lines}
\end{figure*}

\begin{figure*}[h!]
\centering
\includegraphics[scale=0.5]{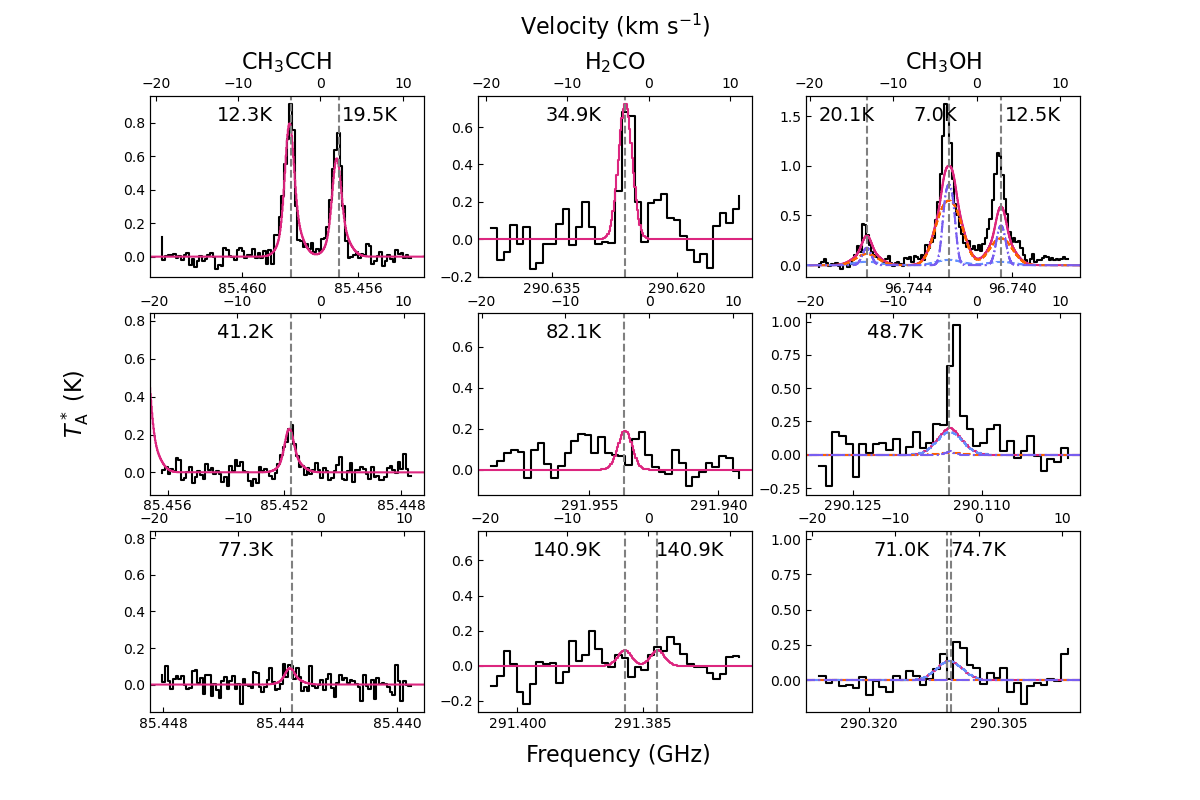}
\caption{Same as Fig.~\ref{fig:n44lines} for the central pixel of N40.}
\label{fig:n40lines}
\end{figure*}

\begin{figure*}[h!]
\centering
\includegraphics[scale=0.5]{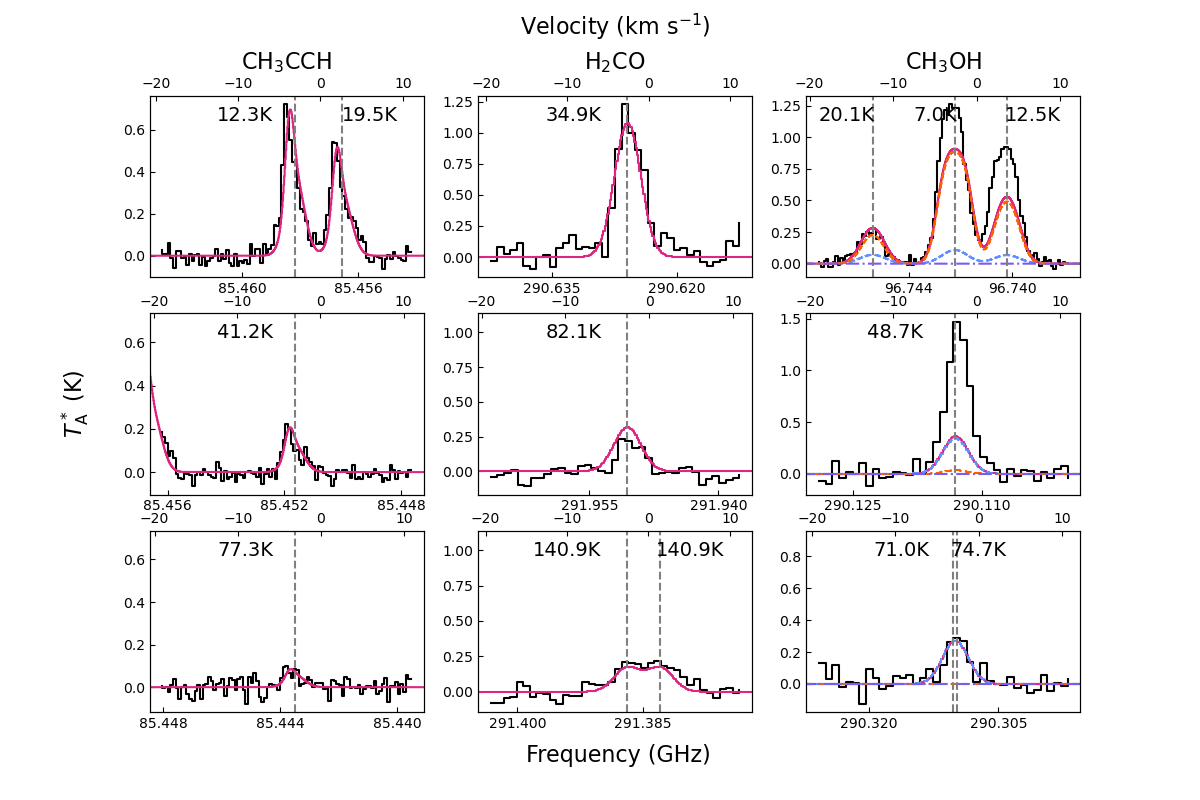}
\caption{Same as Fig.~\ref{fig:n44lines} for the central pixel of N36.}
\label{fig:n36lines}
\end{figure*}

\begin{figure*}[h!]
\centering
\includegraphics[scale=0.5]{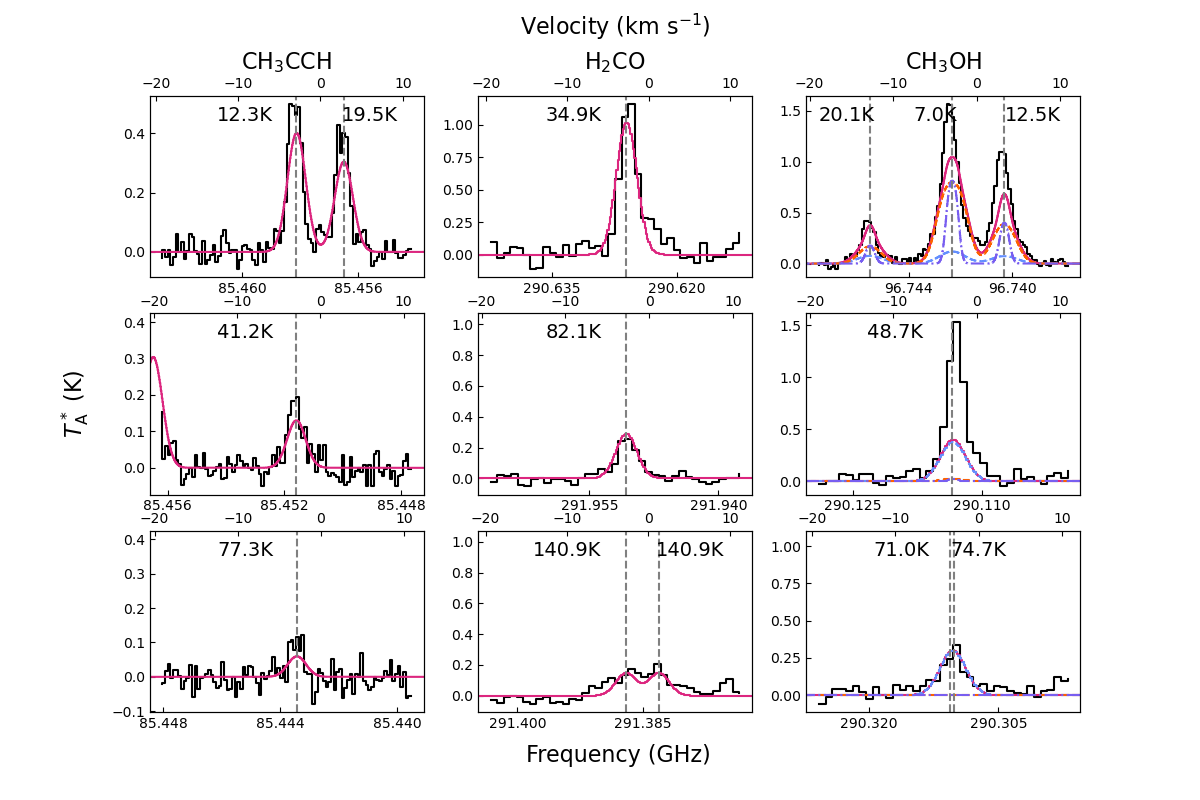}
\caption{Same as Fig.~\ref{fig:n44lines} for the central pixel of N41.}
\label{fig:n41lines}
\end{figure*}

\clearpage

\section{Production and destruction routes from the NAUTILUS model}

\begin{figure*}[h!]
\centering
\includegraphics[trim= 0 1cm 0 2cm, clip=True, scale=0.5]{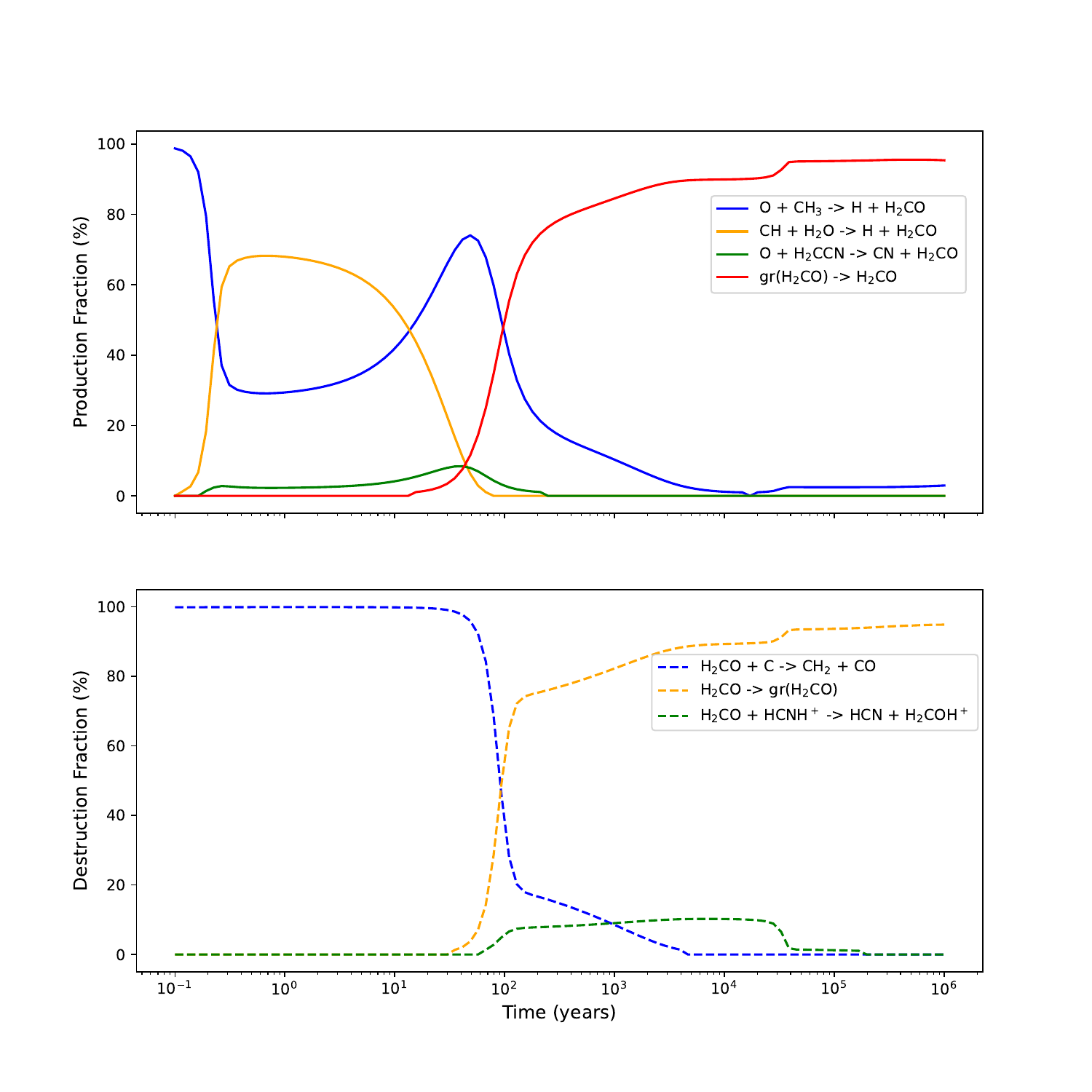}
\caption{The production (top) and destruction (bottom) routes for H$_2$CO, components 1 and 2. The reactions are described in each legend. The designation gr() stands for a grain surface process.}
\label{fig:proddesth2co1}
\end{figure*}

\begin{figure*}[h!]
\centering
\includegraphics[trim= 0 1cm 0 2cm, clip=True, scale=0.5]{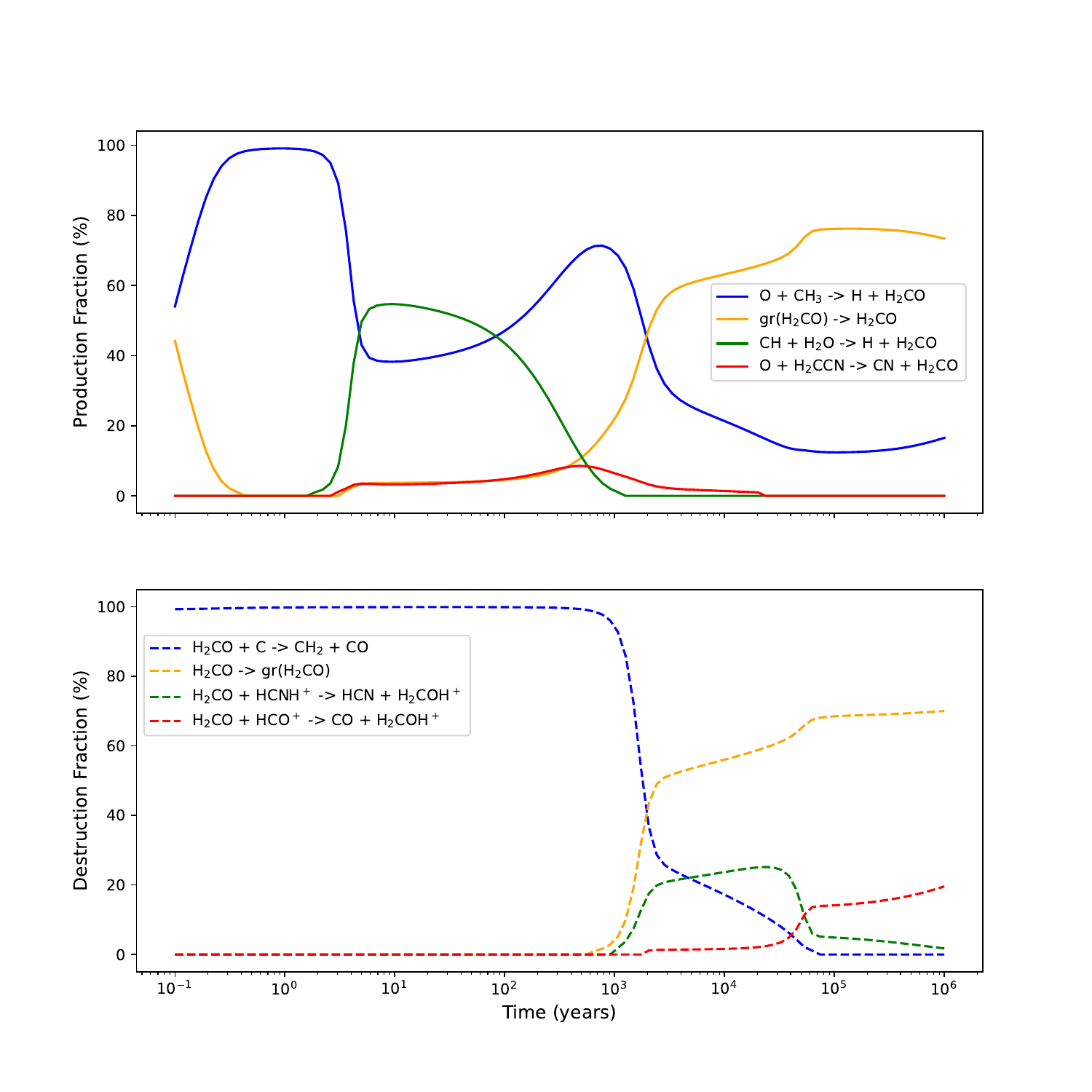}
\caption{Same as Fig.~\ref{fig:proddesth2co1} for H$_2$CO component 3.}
\label{fig:proddesth2co3}
\end{figure*}

\begin{figure*}[h!]
\centering
\includegraphics[trim= 0 1cm 0 2cm, clip=True, scale=0.5]{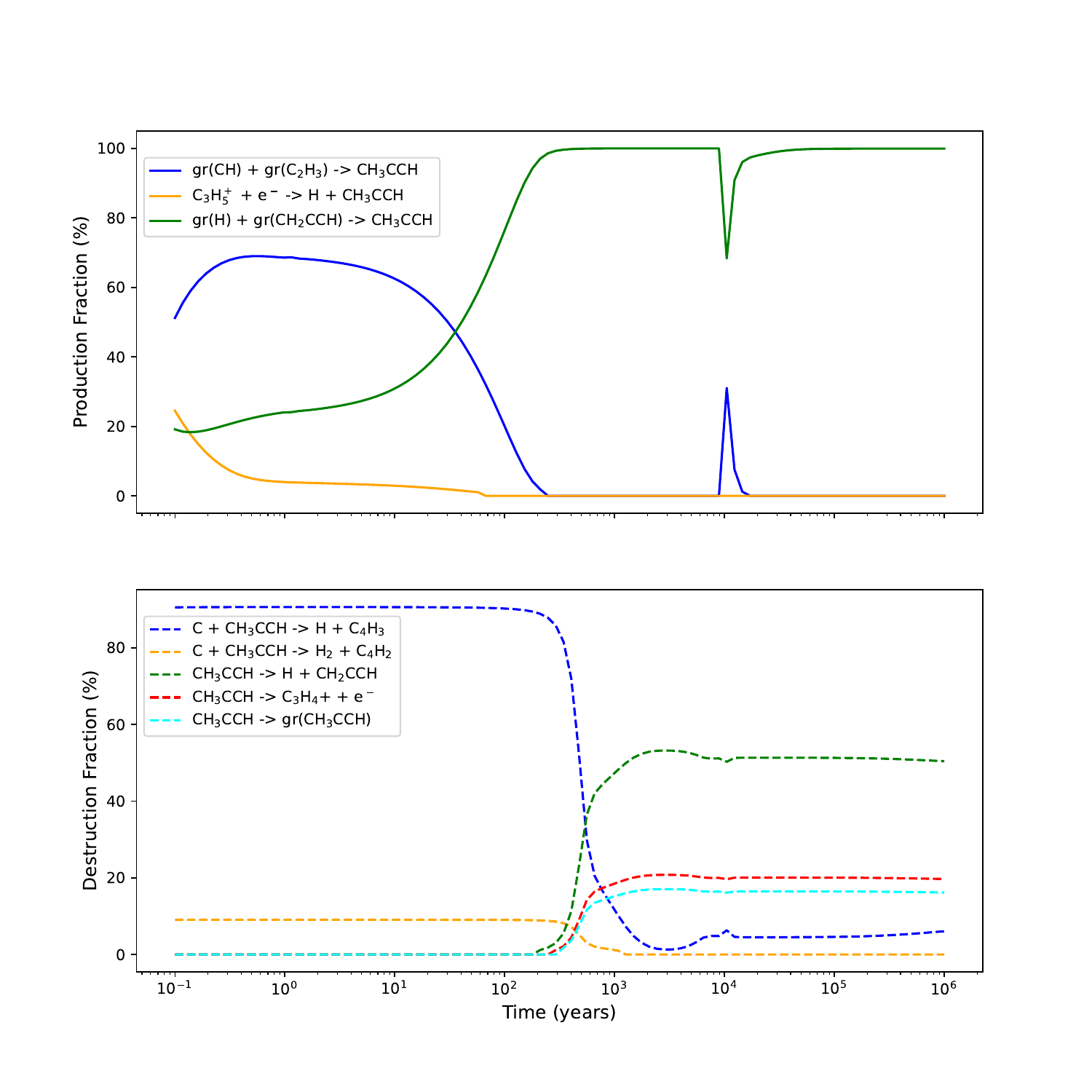}
\caption{Same as Fig.~\ref{fig:proddesth2co1} for CH$_3$CCH, both components.}
\label{fig:proddestch3cch}
\end{figure*}

\begin{figure*}[h!]
\centering
\includegraphics[trim= 0 1cm 0 2cm, clip=True, scale=0.5]{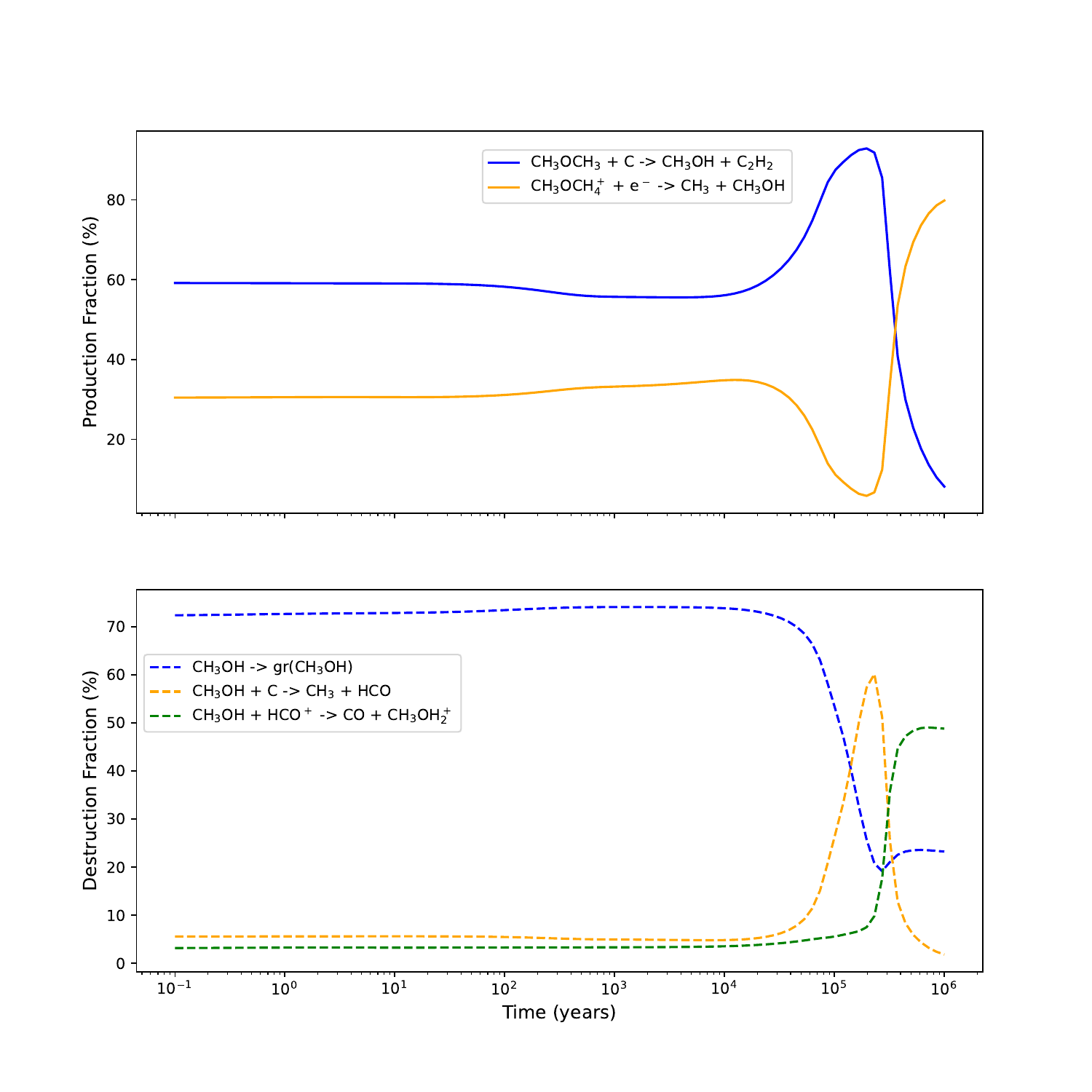}
\caption{Same as Fig.~\ref{fig:proddesth2co1} for CH$_3$OH component 2.}
\label{fig:proddestch3oh2}
\end{figure*}

\begin{figure*}[h!]
\centering
\includegraphics[trim= 0 1cm 0 2cm, clip=True, scale=0.5]{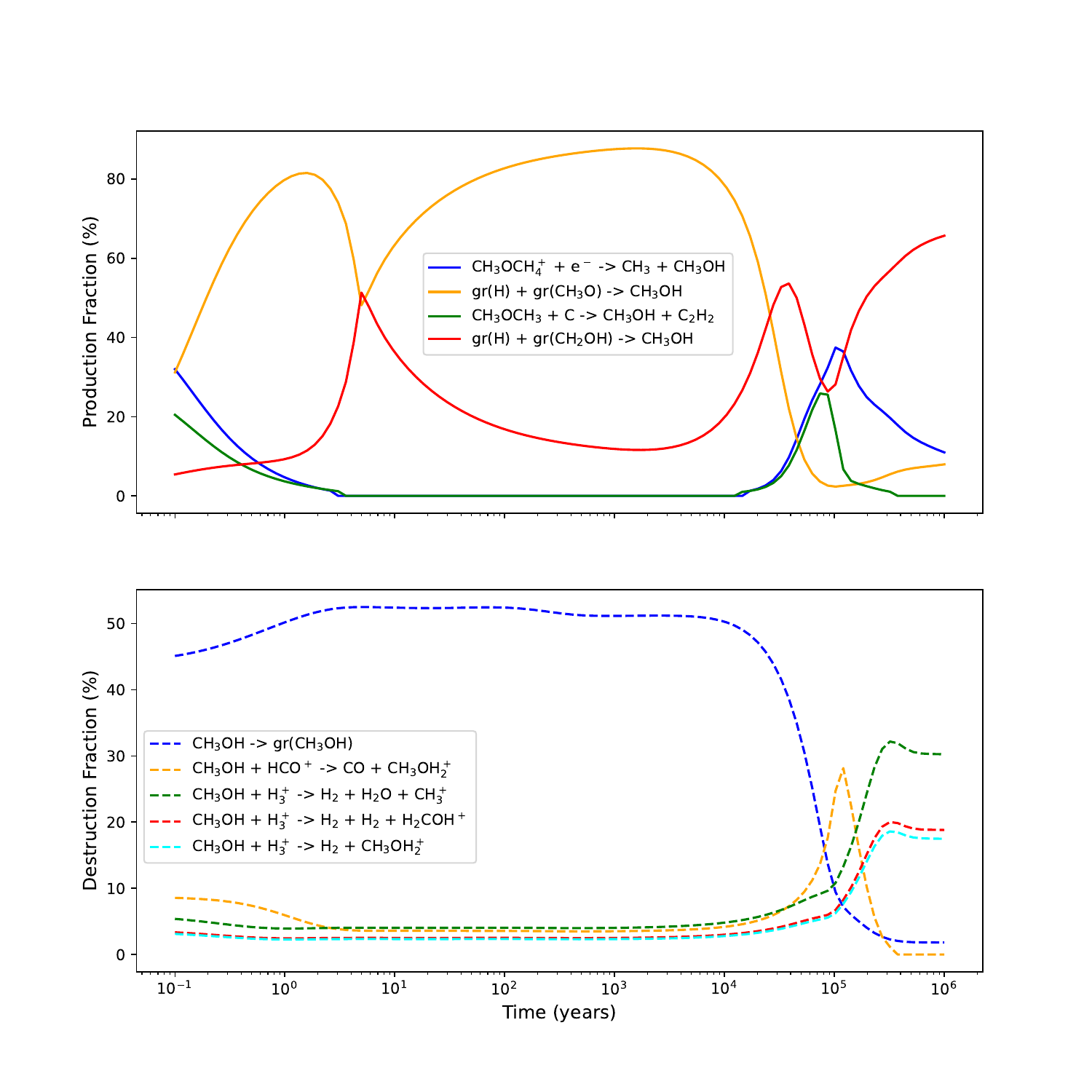}
\caption{Same as Fig.~\ref{fig:proddesth2co1} for CH$_3$OH component 4.}
\label{fig:proddestch3oh4}
\end{figure*}

\begin{figure*}[h!]
\centering
\includegraphics[trim= 0 1cm 0 2cm, clip=True, scale=0.5]{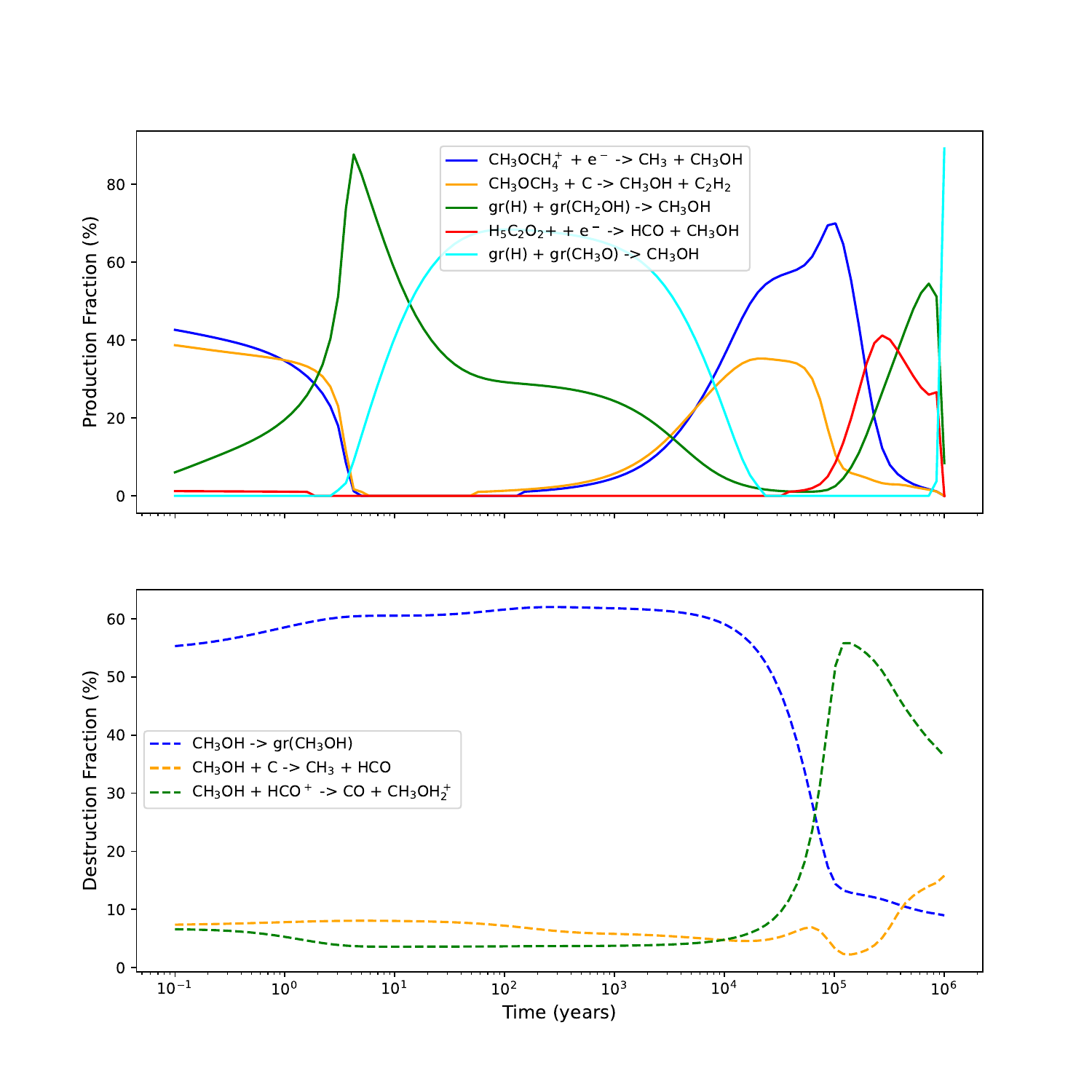}
\caption{Same as Fig.~\ref{fig:proddesth2co1} for CH$_3$OH component 5.}
\label{fig:proddestch3oh5}
\end{figure*}

\end{appendix}

\end{document}